\documentclass[11pt,a4paper]{article}
\usepackage{jheppub,amsmath,  amssymb,slashed,url,bm,textgreek,upgreek,url}
\usepackage{graphicx}
\usepackage{epstopdf}
\usepackage{notoccite}
 
\def\tt{{\mathfrak t}}

\def\be{\begin{equation}}
\def\ee{\end{equation}}

\def\D{{\mathcal D}}

\def\A{{\mathcal A}}
\def\d{{\mathrm d}}

\def\R{{\mathbb R}}

\def\D{{\mathcal D}}
\def\[{\bigl [}

\def\]{\bigr ]}

\def\N{{\mathcal N}}

\def\Z{{\mathbb Z}}

\def\I{{\mathcal I}}

\def\H{{\mathcal H}}

\def\bar{\overline}

\font\teneurm=eurm10 \font\seveneurm=eurm7  \font\fiveeurm=eurm5
\newfam\eurmfam
\textfont\eurmfam=\teneurm \scriptfont\eurmfam=\seveneurm
\scriptscriptfont\eurmfam=\fiveeurm

\font\teneusm=eusm10 \font\seveneusm=eusm7 \font\fiveeusm=eusm5
\newfam\eusmfam
\textfont\eusmfam=\teneusm \scriptfont\eusmfam=\seveneusm
\scriptscriptfont\eusmfam=\fiveeusm

\font\tencmmib=cmmib10 \skewchar\tencmmib='177
\font\sevencmmib=cmmib7 \skewchar\sevencmmib='177
\font\fivecmmib=cmmib5 \skewchar\fivecmmib='177
\newfam\cmmibfam
\textfont\cmmibfam=\tencmmib \scriptfont\cmmibfam=\sevencmmib
\scriptscriptfont\cmmibfam=\fivecmmib

\def\d{{\mathrm d}}
\def\i{{\mathrm i}}
\def\sld{{\slashed{D}}}
\def\la{\langle}
\def\ra{\rangle}
\def\Tr{{\mathrm{Tr}}}
\def\sL{{\sf L}}
\def\APS{{\mathrm{APS}}}
\def\sgn{{\mathrm{sign}}}
\def\1{{\mathbf 1}}

\title{A Note On Lattice Definition of Gauge Theories With Chiral Fermions}
 \author{Edward Witten}
\affiliation{School of Natural Sciences, Institute for Advanced Study,\\ 1 Einstein Drive, Princeton, NJ 08540 USA}
\abstract{Because of the fermion doubling problem, it has been a longstanding challenge to describe a lattice algorithm for studying four-dimensional gauge theories
coupled to chiral fermions.  Recently, an algorithm to do so has been described by Craig, extending partial results in the previous literature.  Here we describe a variant of
this algorithm, based on interpolating a lattice gauge field over the continuum (as in Craig's work) and coupling the interpolated gauge field to continuum chiral fermions.
The same variant was also proposed recently by Hsu.   In any dimension and for any anomaly-free representation,
this approach can be implemented  on a computer in principle, but not in practice by Monte Carlo methods because the chiral fermion path integral is not positive. 
}

\begin{document}\maketitle

\section{Introduction}\label{intro} 

Lattice gauge theory  \cite{Wilson} provides a very satisfactory regularization of four-dimensional Yang-Mills theory, and
has offered an effective starting point for nonperturbative numerical exploration  of this theory, as pioneered in \cite{Creutz}.
The same framework is applicable -- with considerably more difficulty -- for gauge theories like QCD with fermions in vector-like representations.    
However, because of the fermion doubling problem, which is well characterized by the Nielsen-Ninomiya theorem \cite{NielsenNinomiya},
it has been notoriously difficult to describe in this framework  gauge theories with chiral fermions, such as the Standard Model of particle physics.
 
Ginsparg and Wilson proposed a strategy to circumvent the fermion doubling problem \cite{GW}.   Building on the domain wall fermions of Kaplan \cite{Kaplan},
  Narayanan and Neuberger \cite{NN,NN2} interpreted the chiral fermion determinant as an overlap.   Neuberger \cite{N} and Hasenfratz  \cite{H} then derived lattice Dirac operators that satisfy
  the Ginsparg-Wilson relation, and L\"uscher \cite{Luscher} showed that this implies an exact  chiral symmetry of the classical lattice action.
But this framework does
not provide any immediate way to define a suitable measure for integration over these fermions, so quantization is not obvious.
However,   L\"uscher was able to define such a measure for chiral gauge theories with abelian gauge group \cite{Luscher2}, thus achieving a lattice regularization for such theories.   One important ingredient
in this construction was a restriction to ``admissible'' gauge fields -- gauge fields whose plaquette curvatures are small.  

The extension of L\"uscher's construction to nonabelian gauge theory
was not straightforward.   L\"uscher himself showed that a nonabelian version of his construction works to all orders of perturbation theory around the trivial connection \cite{Luscher21},
and there were several other partial results  \cite{S,KK,K}.
Very recently, Craig overcame the difficulties and, with the aid of several new ideas,  finally constructed a lattice regularization of chiral gauge theories in enough generality to encompass
the Standard Model  \cite{Craig}.

In this article, we will revisit this subject, and explain a slightly different route to the objective
using two key facts.   The second of these facts was  also used
by Craig, and the first is closely related to facts that he used.  Actually, following the appearance of Craig's paper and while the present one was being written, another paper appeared by Hsu \cite{Hsu}
also describing a lattice regularization of chiral fermions and using the same key facts that the present article is based on.  We will say more about Hsu's paper and the background to it in section \ref{facttwo}.

Our analysis gives several generalizations of previous results (very likely the arguments in \cite{Craig,Hsu} can be modestly extended to yield these generalizations).
First, we use no special properties of the gauge group or fermion representation, so in four dimensions our analysis applies for any gauge group and anomaly-free fermion representation.
Second (provided one follows the general analysis in section 2.4 of \cite{WY} or in the appendix of \cite{FH} rather than the simpler explanation we give in section \ref{startup}), our discussion generalizes 
to any anomaly-free fermion system in any dimension, such as chiral fermions in two dimensions or more exotic cases such as nonperturbative      anomalies of three-dimensional fermions in theories
with time-reversal symmetry.  Finally,
 our discussion applies on an arbitrary compact Euclidean spin manifold, not necessarily a torus.   One might question the importance of this generalization, as lattice gauge theory is usually
formulated on a torus.   However, since gravity does exist in the real world, part of the consistency of anomaly-free chiral gauge theories such as the Standard Model is that they
can be consistently coupled to gravity. And part of the consistency of that coupling is that these models can be defined in Euclidean signature on any spin manifold.

 We now summarize the two key facts that we will use.   This summary is detailed enough to explain the main ideas of the paper.

\subsection{The  Fermion Path Integral In A Background Field}\label{factone}

The first key fact is that in continuum Yang-Mills theory with any gauge group $G$, the path integral for fermions in any anomaly-free representation,   chiral or not,
coupled to any classical background field $A$  (and any background spacetime metric), is always well-defined.  The representation has to be free of both perturbative and nonperturbative (global) anomalies.
This statement may sound like a tautology; if the fermion path integral in some representation
is not well-defined, we declare that this representation is anomalous.    What gives the statement its force is the following:

(1) In learning how to define the fermion path integral for anomaly-free theories, one also learns to determine all possible nonperturbative anomalies -- that is, anomalies
beyond the standard perturbative  anomalies.   For example, in four-dimensional chiral gauge theories with connected and simply-connected gauge group, one learns \cite{Freed,WY} that
the only anomaly beyond the usual triangle anomalies\footnote{By the usual triangle anomalies, we intend both gauge anomalies and the mixed gauge-gravitational anomaly
that arises in $U(1)$  gauge theory if the sum of the charges of positive chirality fermions does not vanish.   The latter anomaly obstructs the ability to define
a model on a general four-dimensional spin manifold.
 If one wishes in lattice gauge theory to study a $U(1)$ theory that has no gauge anomaly but does have a mixed
gauge-gravitational anomaly, then one has  to choose the spacetime $W$ (and the five-manifold $Y$ introduced later) to be
 manifolds whose first Pontryagin form vanishes, for example products $T^2\times B$, where $T^2$ is a flat two-torus and $B$ is arbitrary.}
is the familiar mod 2 anomaly involving the number of fermion zero-modes in an instanton field \cite{Witten1982,NewSU2}.   Though the Standard Model gauge group is not simply-connected,
the Standard Model fermion representation is completely anomaly-free because it extends to a representation of the simply-connected group $SU(5)$, and this representation is anomaly-free
by the criterion just stated.

(2) The path integral of an anomaly-free fermion system, chiral or not,  can be described by explicit formulas that could be evaluated numerically on a computer.

These statements are valid  in any spacetime dimension
and without assuming spacetime to be orientable (as it need not be in a parity-conserving theory),\footnote{For analysis of a 
general case, 
see section 2.4 of \cite{WY} or the appendix of \cite{FH}.  Once can also generalize from spin structures to spin$_c$ structures, etc.}  but  we will in this
article concentrate on theories defined on oriented  spacetimes $W$ of dimension $d=4$.   $W$ will  be assumed to have a spin structure so that fermions can be defined on $W$.   A generalization beyond the familiar case $W=T^4$ is introduced explicitly only at the end of section \ref{interpolation}.

The positive chirality fermions of a four-dimensional gauge theory, taken together, transform in a possibly reducible representation $R$ of
the gauge group $G$. Let $\sld_W^+$ be the Dirac operator acting on this representation.   Formally the fermion partition function  is\footnote{For the moment, we assume
that instanton numbers vanish so that $\det\sld_W^+$ is generically nonzero.   The generalization is discussed briefly below and in more detail in section \ref{goingon}.}    $\det\sld_W^+$, the determinant of $\sld_W^+$.    There is no problem to define the absolute value $|\det\,\sld_W^+|$ of this determinant;
it is the square root of the determinant of the positive operator $\sld_W^{+\dagger}\sld_W^+$, and can be defined by any standard regularization, such as Pauli-Villars or zeta-function regularization.
To define the phase of $\det\, \sld_W^+$ is subtle.   If, however, $W$ is the boundary of a manifold $Y$ of dimension five over which the gauge field $A$ and the orientation and spin structure of $W$ extend,
then we can proceed as follows.   
Let $\sld_W$ be a {\it non-chiral} Dirac operator acting on fermions that transform in the representation $R$.  $\sld_W$ anticommutes with the usual $\gamma_5$ operator $\gamma_5=\gamma_1\gamma_2\gamma_3\gamma_4$ (which can be defined because $W$ is assumed to be oriented).   Using $\gamma_5$ as a fifth gamma matrix, we can define
a five-dimensional Dirac operator $\D_Y=\i\sld_Y$ that is equivalent to  $\i\sld_W$ along the boundary.   Let $\eta(\D_Y)$ be the Atiyah-Patodi-Singer (APS) eta-invariant of this operator, computed with APS boundary
conditions.   Then, as we will explain, a formula for the fermion
partition function, valid in any anomaly-free representation,  is\footnote{The formula (\ref{matchone}) may appear to have a  non-smooth dependence on $A$ at the locus in field
space along which $\sld_W^+$ develops a zero-mode, causing  $\det\,\sld_W^+$ to vanish.  This is not the case; along this locus, $\eta(\D_Y)$ has a non-smooth behavior such that the right hand side of
eqn. (\ref{matchone}) is perfectly smooth.}
\be\label{matchone}|\det\,\sld_W^+| \exp(-\i\pi \eta(\D_Y)). \ee
Once the fermion partition function is defined (and  a suitable alternative is described  for instanton sectors, where the partition function vanishes), there is no problem  to define arbitrary correlation functions of operators involving fermions.   The only new ingredient required is the fermion propagator (or Green's
function), and this is well-defined for any fermion representation, anomaly-free or not.

As written, the formula (\ref{matchone}) 
appears to depend on the choice of the five-manifold $Y$ and the extension of the gauge field $A$ over $Y$.   However, it turns out that the fermion representation $R$
is anomaly-free if and only if the exponentiated eta-invariant  $\exp(-\i\pi\eta(\D_Y))$  depends only on $W$ and the gauge field, metric, and spin structure of  $W$, not on $Y$ or the extension of that data over $Y$.   This statement
can be motivated  as follows (see section \ref{chiral} for a technical explanation).   First of all,  perturbative anomalies in $d$ dimensions are related to Chern-Simons forms in $d+1$ dimensions \cite{J,Z}.
This relation is beautifully expressed in the idea of anomaly inflow from an extra dimension \cite{CH}.  If and only if a representation $R$ has no perturbative anomaly, the Chern-Simons form  that would be related to that representation by anomaly inflow vanishes.  The eta-invariant of a connection is locally equivalent to the integral of a Chern-Simons form, but globally it
captures some additional information.   It turns out that to extend the relation between anomalies and Chern-Simons to global or nonperturbative anomalies, one just has to use the eta-invariant
instead of Chern-Simons.   This was first shown in
\cite{WittenGlobal}, with an important later generalization by  Dai and Freed \cite{DF}.  
The analysis in \cite{WittenGlobal} was based on computing the holonomy around a loop in field space
of what is nowadays often called the Berry connection, or in this context sometimes the Bismut-Freed connection \cite{BF},
of the fermion ground state.  It was shown that this holonomy can be expressed in terms of the eta-invariant.   See also \cite{ADD} for a closely related analysis. 
The upshot  is the following:
a fermion representation $R$ is completely anomaly-free if and only if, for any manifold $Y$ without boundary, $\exp(-\i\pi\eta(\D_Y))=1$.   When this fact is combined with the gluing law for the eta-invariant
that was proved in \cite{DF}, one learns that the right hand side of eqn. (\ref{matchone}) is independent of the choice of $Y$, and of the extension of $A$ over $Y$, if and only if the representation $R$ is
anomaly-free.   
The formula (\ref{matchone}) was most fully justified from a physical point of view in \cite{WY}.   (The Dai-Freed theorem -- as opposed
to the formula (\ref{matchone}) that it suggests -- was elucidated from a physical point of view in  \cite{Y}.)  
The formula (\ref{matchone}) can be viewed as an extension of the fact that anomaly inflow for domain wall fermions can be described via the eta-invariant \cite{KS}.
This formula  has had a variety of applications, for example to the worldsheet path integral of the heterotic string \cite{WittenDInst}.   

This reasoning as we have presented it does not apply in instanton sectors, where $Y$ does not exist and $\det\,\sld_W^+$ typically is identically zero because of fermion zero-modes.   
Two refinements are needed, as elaborated
in section \ref{goingon}: instead of the formula (\ref{matchone}), one characterizes the fermion measure via a similar formula for correlation functions that are generically nonzero; and instead of $Y$
having boundary $W$, its boundary consists of $W$ and a disjoint union of fiducial manifolds $W_i$ equipped with chosen gauge fields
that carry the appropriate instanton numbers.   By specifying the phases of the path integral measure for the fiducial manifolds and gauge fields,
one specifies the renormalized values of gauge and gravitational theta-angles, in general including discrete ones.

An important consequence of the relation of perturbative anomalies to the Chern-Simons form  and thereby  to the eta-invariant, together with the APS index theorem, is the following.
In any $d$-dimensional theory that is free of perturbative anomalies, the $d+1$-dimensional invariant $\exp(-\i\pi\eta(\D_Y))$ that controls global anomalies is a cobordism invariant.   This typically makes it possible  to identify all possible global anomalies in particular theories or show that they are absent  \cite{WittenOlder}, \cite{WittenOtherOld}, \cite{Freed}, \cite{WY}. 

\subsection{Interpolating A Lattice Gauge Field To The Continuum}\label{facttwo}

Now we come to the second key fact.
It may not be obvious that understanding the path integral for anomaly-free chiral fermions  in the continuum is going to help us construct a nonperturbative regularization of chiral gauge theories.
For that purpose, we presumably need to consider {\it lattice} gauge fields.  What good will it do to know what happens if chiral fermions are coupled to {\it continuum} gauge fields?
The answer to this question is clear from the construction described by Craig \cite{Craig} and the further discussion by Hsu \cite{Hsu}. 

By way of background, long ago L\"uscher \cite{Luscher3} introduced the idea of constraining lattice gauge fields so that every
plaquette holonomy $U_P$ is close to the identity, say $|U_P-1|<\epsilon$ for some fixed and sufficiently small $\epsilon$.  Lattice gauge fields with this property
have been called ``admissible.''   The original goal was to define topological invariants such as instanton number in lattice
gauge theory.   In conventional lattice gauge theory, with the link variables being general gauge group elements as in \cite{Wilson}, it is not possible to define a topological invariant such as the instanton number,
because an instanton can continuously shrink to a point and disappear in the holes between the lattice sites.     However,  if one constrains the lattice gauge field so that all plaquette holonomies
are sufficiently close to the identity, then the instanton number (and corresponding topological invariants in other dimensions) can indeed be defined.

It was further argued \cite{Luscher3} that restricting to admissible lattice gauge fields does not affect the continuum limit of the theory.   The reason for this
expectation is that if $g$ is the gauge coupling at the lattice scale, then typically $|U_P-1|\sim g$, which because of asymptotic freedom becomes much less than $\epsilon$ near the continuum limit.
The admissibility constraint is hence expected to become irrelevant in the continuum limit.

Another interpretation of the fact that  instanton number can be defined for admissible gauge fields  
is that admissible lattice gauge fields can be interpolated over the continuum, as shown in various degrees of generality in \cite{GKSW,PS2,Hooft,HS} and \cite{Craig}.   In other words, given an admissible lattice 
gauge field, one can interpolate it to a corresponding continuum gauge field that agrees with the lattice gauge field on the lattice links.
To make 
such an interpolation involves  arbitrary choices, but the choices  can be made in a way that respects all lattice symmetries and is gauge-invariant  (if two lattice gauge fields are gauge-equivalent, then the corresponding
interpolations are also gauge-equivalent) and in such a way that the continuum gauge field associated to a given admissible lattice gauge field always
has field strength (in lattice units) at most of order $\epsilon$.   Under this condition, the topology of the continuum gauge field does not depend on the choices.  We will refer
to such a procedure as  continuum interpolation of admissible lattice gauge fields.   
  Once one knows that  an admissible lattice gauge field has a continuum interpolation, one can simply define its topological invariants to be those of the interpolation.
  
Continuum interpolations exist for every compact gauge group $G$ and on any manifold of any dimension $d$.
We give one explanation of this in section \ref{interpolation}; this can also be deduced by a slight extension of arguments of 't Hooft \cite{Hooft} and Craig \cite{Craig},
among others.

Craig makes use of the existence of continuum interpolations to define a regularization of chiral gauge theories as follows. As explained earlier, overlap fermions and the Ginsparg-Wilson equation
give a way to define chiral fermions on a lattice, but without a simple way to define a measure for integrating over them.    To determine such a measure, Craig
compares chiral fermions on ever finer lattices, ultimately defining the fermion measure via a continuum limit. The continuum interpolation of the original lattice gauge fields is used to
define a gauge field that can be coupled to the lattice fermions on any scale.  In short, Craig uses the behavior near a continuum limit to define a measure for lattice chiral fermions.
   In part, this construction involves new {\it ab initio} proofs of some of the facts
that were cited in section \ref{factone} to explain the well-definedness of the chiral fermion path integral.

Alternatively, and as described independently of the present work 
by Hsu \cite{Hsu}, once one knows that lattice gauge theory can be restricted to admissible lattice gauge fields and that admissible lattice gauge fields have a continuum interpolation, one can construct
a lattice regularization of chiral gauge theories by  directly using the fact that the continuum path integral of anomaly-free chiral fermions is always well-defined.  
In fact, there have been a number of earlier proposals along roughly these lines, including  \cite{FW,GS,Hsu2,Kronfeld,Montvay,HS2} and \cite{Hooft}.    The main novelty of the present article is possibly to incorporate a more precise description of the chiral fermion path integral and its anomalies.

The approach we discuss applies in any
spacetime dimension, for any anomaly-free fermion representation; the only lattice specific fact that is used is the existence of continuum interpolations.
One simply couples admissible lattice gauge fields, or more precisely their continuum interpolation, to  continuum chiral fermions.   There is no fermion doubling problem because one never introduces
lattice fermions (as explained in section \ref{effective}, numerical evaluation does not bring back the doubling problem).
The chiral fermion path integral  is then characterized  by 
 the formula (\ref{matchone}) (or its generalization, explained in section \ref{goingon}, in instanton sectors).    
Thus the lattice path integral is defined by an integral over admissible families of link variables  $U_{\langle xy\rangle}$ connecting neighboring lattice sites
$x,y$, weighted by the exponential of the lattice action $I$ times the output of the fermion path integral.   The instanton number zero contribution to the partition function is 
\be\label{momo} Z= \int_{\mathcal A}\prod_{\langle xy\rangle} \d U_{\langle xy\rangle} \,e^{-I} \,|\det\,\sld_W^+| \exp(-\i\pi \eta(\D_Y)). \ee
Here the integral runs over the space ${\mathcal A}$ of admissible lattice gauge fields, and  the Dirac operators $\sld_W^+$ and $\D_Y$ are constructed using the  interpolated continuum gauge field on $W$
and an extension of it over $Y$.
Of course, we want to calculate not just a partition function but  correlation functions 
of gauge-invariant operators.   These operators are constructed in the usual way as functions of the link variables $U_{\langle xy\rangle}$ along with  the continuum fermion fields evaluated at lattice points.
From any product of such operators, the fermions can be integrated out at fixed values of the $U_{\langle xy\rangle}$, using their propagators (in the  interpolated continuum gauge field).
Then the desired expectation value is evaluated by integrating over the link variables and dividing by the partition function.    Contributions to the path integral from sectors with nonzero instanton numbers are evaluated
by replacing eqn. (\ref{matchone}) with its appropriate generalization, as detailed in section \ref{goingon}.

\subsection{Organization of This Article}\label{organize}

In section \ref{chiral} of this article, we will provide more detail on the well-definedness of the path integral for anomaly-free chiral fermions coupled to a background gauge field.
In section \ref{interpolation}, we explain one approach to constructing continuum interpolations of admissible lattice gauge fields, for an arbitrary compact gauge group $G$ and on a manifold of any dimension.
Finally, in section \ref{effective}, we address the following question:  Could the regularization of chiral gauge theories described here be implemented on a computer in principle, even if not
very efficiently?  One may worry  that when one tries to actually evaluate numerically the formula for the chiral fermion path integral that we describe, the fermion doubling problem
would come back.   We will see that that is not the case.

\section{The Path Integral of Anomaly-Free Chiral Fermions}\label{chiral}

The action for chiral fermions in four-dimensions (on a manifold $W$ of  Euclidean signature) has the general form
\be\label{chirac}I=-\int \d^4x\sqrt g \bar\Psi \slashed{D}_W \Psi,\ee
where we collect all positive chirality fermions in $\Psi$, 
and negative chirality fermions in $\bar\Psi$. $\Psi$ transforms in some possibly reducible representation $R$ of the gauge group, and $\bar\Psi$ transforms in the complex conjugate representation
$\bar R$.    Classically, such a theory has an  evident $U(1)$ symmetry under which $\Psi$ and $\bar\Psi$
 respectively have  charge 1 and $-1$.   This symmetry is anomalous, but it can still be used to organize an analysis
 of the fermion path integral. As in sections 2.1-2.3 of \cite{WY}, exploiting such a symmetry leads to a simple explanation of the relation between anomaly inflow and the eta-invariant.   A completely general
 explanation, applicable for example to chiral fermions in two dimensions and to global anomalies of fermions in three dimensions, can be found in \cite{FH} or section 2.4 of \cite{WY}.
 The main difference is that in the absence of a convenient $U(1)$ symmetry, a more involved notation is required.
 
 In section \ref{startup}, we explain the basic anomaly inflow formula (\ref{matchone}) for chiral gauge theory on a compact spin manifold $W$ of Euclidean signature, assuming
 that $W$ is the boundary of a five-manifold $Y$ over which all relevant structures extend.   This in particular is only true if all relevant instanton numbers vanish.  
 In section \ref{anomaly}, we explain how the anomaly inflow formula can be used to effectively classify the possible nonperturbative or global anomalies for a given gauge group and in particular
 to show that the Standard Model is free of such anomalies.    In section \ref{goingon},
 we explain the generalization when $Y$ does not exist.   The main idea of the generalization is that in defining the theory, for every generator of the cobordism group that obstructs the existence
 of $Y$, one has to introduce a dual topological theta-angle (discrete or continuous depending on whether the generator in question is a torsion element or not).
 
 \subsection{Anomaly Inflow and the Eta-Invariant}\label{startup}
 
To study the chiral fermion theory (\ref{chirac}),  let $\Psi$ be a {\it non-chiral} fermion field in the representation $R$.
 Then, on an oriented four-manifold $W$, we can define  the chirality operator  $\gamma_5=\gamma_1\gamma_2\gamma_3\gamma_4$, and using this as the fifth gamma matrix, we can define a five-dimensional
 Dirac equation for a fermion in the representation $R$.   This five-dimensional fermion can have a bare mass, the action  on an oriented five-manifold $Y$ being
 \be\label{hirac} I= -\int_Y\d^5x \sqrt g \bar\Psi\left(\slashed{D}_Y+m\right)\Psi . \ee  
 To write this action, we added a five-dimensional fermion field $\bar\Psi$ in the complex conjugate representation $\bar R$.
 This theory is completely anomaly-free, since a bare mass is possible.   For example, a Pauli-Villars regulator is possible.  This entails including a field obeying the same Dirac equation as $\Psi$
 but with opposite statistics.   Roughly speaking, the regularized fermion path integral is then  \be\label{tirac}Z_\Psi =\prod_k \frac{-\i\lambda_k+m}{-\i\lambda_k+M},\ee
 where $M\gg |m|$ is the  regulator mass, and the product runs over the eigenvalues $\lambda_k$ of the self-adjoint operator $\D_Y=\i\sld_Y$. (If $\lambda_k$ is an eigenvalue of $\D_Y$, then
 the corresponding eigenvalue of $\slashed{D}_Y$ is $-\i\lambda_k$, so $\det(\slashed{D}_Y+m)$ is a product of factors $-i\lambda_k+m$.)   Actually, in five dimensions, the  infinite product in (\ref{tirac}) does not converge.
 One can achieve convergence, for example, by including additional regulator pairs of opposite statistics, thus including a factor $\prod_k\prod_{r=1}^p \frac{-\i\lambda_k+M_r}{-\i\lambda_k+M'_r}$ with
 suitable regulator masses $M_r$, $M_r'$. (One also should multiply the expression for $Z_\Psi$ by the exponential of a local counterterm to remove scheme-dependent local contributions to the effective action.)
For brevity, we generally omit such additional regulators and describe  Pauli-Villars regularization by the simplified formula (\ref{tirac}).
 
 Now suppose that $Y$ has boundary $W$, and that near $W$, the metric of $Y$ has the form $\d s^2=\d\tau^2+\d s_W^2$, where the boundary is at $\tau=0$, the normal direction to the boundary
 is parametrized by $\tau<0$, and $\d s^2_W$ is some metric on $W$.
 We assume that this form of the metric is valid in  an interval $\tau\in (-\epsilon,0]$.   We impose on $\Psi$ and $\bar\Psi$ a local boundary condition along $W$:
 \be\label{locb} \sL: \left.(1-\gamma^\tau)\Psi\right|_{\tau=0}=0 =\left.\bar\Psi(1+\gamma^\tau)\right|_{\tau=0},\ee
 where $\gamma^\tau$ is the gamma matrix in the $\tau$ direction.\footnote{The opposite signs in the boundary conditions for $\Psi$ and $\bar\Psi$ ensure that
 the boundary term $\int_W\d^4x\sqrt g\bar\Psi \gamma^\tau\delta\Psi$ in the variation of the action vanishes, so that the variational problem is well-posed and $\Psi$ and $\bar\Psi$
 each obey simple Dirac equations with no boundary correction.}    
 This boundary condition is elliptic but not self-adjoint. Ellipticity alone does not ensure a discrete spectrum in finite volume
 (for example, every complex number is an eigenvalue of the elliptic operator $e^{i\theta} \frac{\d}{\d\theta}$
  on a circle), but here discreteness can be shown as follows.. Integrating by parts, one finds that if $\psi$  satisfies the boundary condition
  $\left.(1 -\gamma^\tau)\psi\right\vert_W = 0$, then $2\, {\mathrm {Re}}\int_Y \d^5x\sqrt g
  \bar\Psi  \sld_Y \Psi = \int_W \d^4x \sqrt g |\Psi|^2 \geq 0$. Hence $\sld_Y + m$ has no kernel for ${\mathrm {Re}}\, m > 0$, and a similar argument for the adjoint operator, whose boundary condition is $\left.(1 +\gamma^\tau)\Phi\right|_W = 0$, shows that $\sld_Y$ is  invertible. Its inverse is compact by elliptic regularity, so $\i\sld_Y$ with this boundary condition has a discrete spectrum. The eigenvalues of $\i\sld_Y$
  are not real.  If they were real, the chiral physics along the boundary that we are about to describe could not arise.  But the discrete spectrum makes it possible to define the fermion determinant and propagator and to do
  quantum field theory with this boundary condition.
For a Lorentz signature  explanation of why the boundary condition is physically sensible,  see \cite{WY}.
 
 The action (\ref{hirac}) can be written near the boundary as
 \be\label{zirac} I=-\int_Y \d^4x \d\tau \sqrt g \bar\Psi \gamma^\tau\left(\frac{\partial}{\partial\tau}+\D_W+\gamma^\tau m\right)\Psi\ee
 where
 \be\label{nirac}\D_W=\sum_{\mu\not=\tau}\gamma^\tau\gamma^\mu D_\mu. \ee
 $\D_W$ is equivalent to the usual Dirac operator $\i\sld_W$, differing from it only by the use of a different representation of the same Clifford algebra (with $\gamma^\mu$ replaced by $-\i\gamma^\tau\gamma^\mu$).
 
 So far, $m$ could be positive or negative, but given our choice of boundary condition, the interesting case is $m<0$.   For $m<0$, the Dirac equation $(\sld_Y+m)\Psi$ has solutions  localized near the boundary
of the form
 \be\label{poly} \Psi=\chi \exp(|m|\tau),~~~ (1-\gamma^\tau)\chi=0, ~~~~\D_W\chi=0.\ee
As this solution is localized near $\tau=0$, and  $\chi$ satisfies the four-dimensional Dirac equation $\D_W\chi=0$ along with the four-dimensional chirality condition $(1-\gamma^\tau)\chi=0$, $\chi$ propagates
 along $W$ as a massless chiral fermion.   For $m>0$, there is no such boundary-localized mode.
 
 It is likewise true that for $m<0$, $\bar\Psi$ has a boundary localized mode of the opposite chirality:\footnote{We have converted $\bar\Psi$ and $\bar\chi$ to column spinors by taking their transposes
 and multiplying by a  charge conjugation matrix $C$ that satisfies  $C\gamma_\mu C^{-1}=\gamma_\mu^{\mathrm{tr}}$ (where $\mathrm{tr}$ denotes the transpose).}
 \be\label{noly}\bar\Psi=\bar\chi \exp(|m|\tau),~~~(1+\gamma^\tau)\bar\chi=0,~~~~\D_W\bar\chi=0. \ee
 Considering modes of both $\Psi$ and $\bar\Psi$, the boundary localized fermions are the same ones as in the purely four-dimensional theory (\ref{chirac}): positive chirality fermions in the representation $R$ and negative chirality fermions in the complex conjugate
 representation $\bar R$.   
 
 Depending on $R$, the purely four-dimensional theory may be anomalous, but the five-dimensional theory that generates the same massless fermions along the boundary
 is not anomalous.   In fact, the theory (\ref{hirac}) with  boundary condition $\sL$ is completely anomaly-free.
 One way to see this is to note that if $m$ is positive, there are no light modes at all, and fermion theories with no light modes are always anomaly-free. On the other hand, anomalies never depend on fermion
 bare masses, so the theory must remain anomaly-free for $m<0$.   Even more explicitly, the boundary condition does
 not obstruct the ability to carry out Pauli-Villars regularization.   Finally, this theory actually has a lattice regularization.   Indeed, a lattice regularization of this theory is Shamir's variant \cite{YS} 
 of domain wall fermions \cite{Kaplan};  five-dimensional fermions (which have no lattice doubling problem) on a lattice with boundary, with a suitable ``free'' boundary condition in the lattice theory, can
 lead in a continuum limit to the boundary condition that we have called $\sL$.
 
 As in \cite{WY}, we want to compute the partition function of the massive fermion $\Psi$ with boundary condition $\sL$.   For this, we think of $\tau$ as a Euclidean time coordinate, and the boundary $W=\partial Y$
 as a time-slice.     The path integral over $Y$ then gives a physical state vector that we denote as $|Y\ra$.   This is a state vector in the Hilbert space $\H_W$ of the fermion theory quantized on the spatial 
 manifold $W$.   The boundary condition $\sL$ also corresponds to a state vector $|\sL\ra\in \H_W$, which we might call the boundary state.   The partition function on $Y$ with the boundary condition $\sL$
 is then given by
 \be\label{polme}Z(Y,\sL)=\la\sL|Y\ra. \ee
 
 Now let us take the mass $|m|$ to be very large so that the fermion Compton wavelength is much shorter than the typical scale of manifolds $W$ and $Y$; indeed, we can take $|m|$ all the way up to the
 cutoff, with $m=-M$.  Then the Hilbert space $\H_W$ has a large mass gap from the ground state. Euclidean propagation in the cylindrical region $(-\epsilon,0]\times W$ can be described by
 the action of the operator
 $e^{-\epsilon H}$, where $H$ is the Hamiltonian.   For $\epsilon|m|\gg 1$, $e^{-\epsilon H}$ converges to the projection operator $|\Omega\ra\la\Omega|$ to the ground state $|\Omega\ra$.
 So $|Y\ra$ is a multiple of $|\Omega\ra$, and the formula (\ref{polme}) can be factored
 \be\label{zolme} Z(Y,\sL) =\la\sL|\Omega\ra \la\Omega| Y\ra, \ee
 splitting off the boundary contribution $\la\sL|\Omega\ra$ from the bulk contribution $\la \Omega|Y\ra$.   
 
 In a generic quantum system, there is no natural way to define the phase of the ground state $|\Omega\ra$.   As one varies background fields, $|\Omega\ra$ varies as a section of a line bundle
 over the field space, and one will potentially run into the Berry connection on this line bundle.  Thus, by generic logic, one might expect
 a phase ambiguity in each of the two factors $\la\sL|\Omega\ra $ and $\la\Omega| Y\ra$, and one might anticipate that only their product in eqn. (\ref{zolme}) is well-defined.
 
 Perhaps surprisingly, in the case of free fermions in a background field, this phase ambiguity can be eliminated by making use of the global Atiyah-Patodi-Singer (APS) boundary condition \cite{APS,APS2}.
 Assuming $\D_W$ has no zero-modes (for the generalization see section \ref{goingon}), APS boundary conditions are defined by requiring that $\Psi$, restricted to $W=\partial Y$, is a linear combination
 of eigenfunctions of $\D_W$ with positive eigenvalue.  
  The APS boundary condition is not local, but it is self-adjoint, so  with this boundary condition, $\D_Y=\i\sld_Y$  has a complete set of real eigenvalues.
   Like $\sL$, APS boundary conditions define a boundary state $|\APS\ra$.  Because this state is annihilated by modes of $\Psi$ corresponding to negative eigenvalues, it is a multiple of the state
 $|\Omega_{m=0}\ra$ that would be the fermion ground state if $m=0$.   This state is certainly not the fermion ground state for the case $m<0$ that we are actually studying, but it can be
 viewed (after regularizing high energy modes) as a state in the same Hilbert space.   
 
 Unlike the ground state of a generic quantum system, the state $|\APS\ra$ has no phase ambiguity.   One way to explain that is simply to observe that the inner product $\la\APS|Y\ra$
 is simply the fermion path integral on $Y$ computed with APS boundary conditions.   This is completely well-defined, since it can be defined as in eqn. (\ref{tirac}) using Pauli-Villars regularization,
 but now using the eigenvalues of $\i\sld_Y$ with APS boundary conditions:
 \be\label{zongo}Z_{\Psi,\APS}=\prod_k \frac{-\i\lambda_k+m}{-\i\lambda_k+M} \ee
 (as before, this is a somewhat symbolic formula in which we omit additional regulators).   We will evaluate this expression in a moment.
 
 Using the fact that $|Y\ra$ is a multiple of $|\Omega\ra$, we  now can write an improved version of the formula (\ref{zolme}):
 \be\label{bolme} Z(Y,\sL)= \frac{\la\sL|\Omega\ra\la\Omega|\APS\ra}{|\la\APS|\Omega\ra|^2}\cdot \la \APS|Y\ra. \ee   Clearly, this formula is equivalent to (\ref{zolme}) as long as $|Y\ra$ is a multiple of
 $|\Omega\ra$.
 The first factor $\frac{\la\sL|\Omega\ra\la\Omega|\APS\ra}{|\la\APS|\Omega\ra|^2}$ in (\ref{bolme}) knows about the chosen boundary condition $\sL$ but has no knowledge of the manifold $Y$,
 and the second factor $\la\APS|Y\ra$ knows about $Y$ but not about the choice of boundary condition $\sL$.     Since $Z(Y,\sL)$ is well-defined, and the second factor on the right hand side of eqn. (\ref{bolme})
 is also well-defined, 
 evidently so is their ratio, the first factor on the right hand side.  
 
 For free fermions in a background field, the ratio  $\frac{\la\sL|\Omega\ra\la\Omega|\APS\ra}{|\la\APS|\Omega\ra|^2}$ can be evaluated by expanding the field $\Psi$
 in eigenmodes of the Dirac operator $\D_W$ and evaluating the matrix elements on a mode by mode basis, leading to an answer that is a product over
 modes.  The derivation can be found in section 2.3 of \cite{WY}.
 The result\footnote{\label{lookback} To be precise, eqn. (2.31) of \cite{WY} asserts that  $\frac{\la\sL|\Omega\ra\la\Omega|\APS\ra}{|\la\APS|\Omega\ra|^2}$   
 is a regularized version of $\prod_k\frac{|\lambda_k|}{2|m|}$, where the eigenvalues of $\D_W^{+\dagger}\D_W^+$ are $\lambda_k^2$.   After regularization and renormalization,
 the  infinite product of factors $\frac{1}{2|m|}$, one for each mode, disappears
 (its regularized version, for instance using $\zeta$-functions, renormalizes the local terms in the effective action), 
 and  $\prod_k |\lambda_k|$  becomes a regularized version of $|\det\,\D_W^+|$.} is that this quantity is a regularized version of
 the absolute value of the naive path integral $\det\,\D_W^+$ of the chiral fermions that propagate along $W=\partial Y$:
 \be\label{factorone} \frac{\la\sL|\Omega\ra\la\Omega|\APS\ra}{|\la\APS|\Omega\ra|^2}=|\det\,\D_W^+|. \ee
 We will not repeat  here the derivation, which is reminiscent of lattice calculations with overlap
 fermions, but we note the following.   Since chiral fermions  in the representation $R$ do propagate along $W$, their path integral must somehow appear in $Z(Y,\sL)$.   On the other hand,
 the naive path integral $\det\,\D_W^+$ is not well-defined for a generic representation; its absolute value is well-defined, as it is the square root of the determinant of the self-adjoint operator
 $\D_W^{+\dagger} \D_W^+$, but the phase of $\det\,\D_W^+$ is anomalous.    Accordingly, what appears in eqn. (\ref{factorone}) is the well-defined absolute value $|\det\,\D_W^+|$, not
 the ill-defined $\det\,\D_W^+$.   
 
 A point worth noting is that although $|\det\,\D_W^+|$ is well-defined for any metric $g$ and gauge field $A$ on $W$, it fails to be real analytic along the locus
 in field space at which a zero-mode appears and $|\det\,\D_W^+|=0$.   In the context of the formula (\ref{factorone}), this behavior is possible because the APS boundary
 condition, which appears on the left hand side of the formula, fails to be self-adjoint when the operator $\D_W^+$ has a zero-mode, and  the real analyticity of the left
 hand side of the formula also fails along that locus.  
 Likewise, the second factor in eqn. (\ref{bolme}), though gauge-invariant and real analytic
 as a function of $g$ and $A$ as long as $\D_W$ has no zero-modes, becomes ill-defined (and certainly not real analytic) along the  same locus in field space, again because of the breakdown
 of APS boundary conditions.    This does not affect the validity of the formula (\ref{bolme}), because the first factor on the right hand side of
 that formula vanishes when the second factor is ill-defined.   The product $Z(Y,\sL)$ is completely gauge-invariant and real analytic, but the factorization of $Z(Y,\sL)$ in eqn. (\ref{bolme})  breaks down along the locus in field space where $Z(Y,\sL)=0$.
 
Now let us discuss the second factor $\la \APS|Y\ra$.   This is the path integral of a massive fermion on $Y$, with the self-adjoint APS boundary condition.  It has long been understood
that integrating out a massive fermion on an odd-dimensional manifold without boundary produces a Chern-Simons function perturbatively \cite{Redlich} and an eta-invariant nonperturbatively
\cite{ADM}.   In the present context, on a manifold with boundary but with the self-adjoint APS boundary condition, the same logic applies; we just have to define the eta-invariant using the
eigenvalues of $\i\slashed{D}_Y$ computed with APS boundary conditions.   We will sketch the derivation, essentially following \cite{ADM}.   We return to the formula (\ref{zongo})
for $Z_{\Psi,\APS}$, where we understand that this formula will require some further regularization.   We are interested in evaluating this formula for the case that $m$ is negative and comparable to
the cutoff $M$.  It is convenient then to set $m=-M$.   But first, we note that if $m=+M$, then $Z_{\Psi,\APS}=1$.   The meaning of this is simply that if $m=+M$, then all modes of $\Psi$ are near the cutoff,
and the partition function obtained by integrating over $\Psi$ should equal 1, after removing inessential scheme-dependent local terms in the effective action.     For $m=-M$, it is evident from the formula
(\ref{zongo}) 
that $|Z_{\Psi,\APS}|=1$, so only the phase of $Z_{\Psi,\APS}$ will be of interest.     For this phase, we can write, at general $m$,
\be\label{argz} {\mathrm {Arg}}(Z_{\Psi,\APS}) =\sum_k {\mathrm {Arg}}\,\frac{-\i\lambda_k+m}{-\i\lambda_k+M}. \ee
Assuming that $|\lambda_k|\ll M$ and $\lambda_k\not=0$, as  $m$ varies along the real axis from $M$ to $-M$,    ${\mathrm {Arg}}\,\frac{-\i\lambda_k+m}{-\i\lambda_k+M}$ varies from $0$ to $-\pi \,\sgn(\lambda_k)$.
Naively then, $ {\mathrm {Arg}}(Z_{\Psi,\APS}) $  should equal $-\pi \,\sum_k \sgn(\lambda_k)$.   However, the formula $-\pi \,\sgn(\lambda_k)$ is only valid if $\lambda_k\ll M$, and we need to cut off
the contribution of very large eigenvalues of $\i\slashed{D}_Y$.   The exact cutoff does not matter, as long as it is smooth and sufficiently close to treating positive and negative $\lambda_k$ symmetrically.
Though we could include additional Pauli-Villars regulator fields, it is more simple to define
\be\label{neta}\eta=\lim_{\epsilon\to 0}\sum_k e^{-\epsilon|\lambda_k|} \sgn(\lambda_k). \ee
The cutoff actually used by Atiyah, Patodi, and Singer was different but  equivalent.   They defined\footnote{We note two points of terminology.   Atiyah, Patodi, and Singer defined $\eta$
by a sum as in (\ref{nbeta}) restricted to $\lambda_k\not=0$.   In a moment, we will depart from this and include the zero-modes in the definition of $\eta$; what we call $\eta$ was called $2\xi$ by Atiyah, Patodi, and Singer.
Also, what we here call $\eta$ was called $\eta_{\mathrm D}$ in \cite{WY}.   In that paper, the name $\eta$ was used for an eta-invariant 
that is defined by summing over modes of both $\Psi$ and $\bar\Psi$ (this definition has a better generalization in $d$ dimensions), and an eta-invariant defined as we do here summing only over modes of $\Psi$ was called $\eta_{\mathrm D}=\tfrac{1}{2}\eta$.}
\be\label{nbeta}\eta=\lim_{s\to 0} \sum_k |\lambda_k|^{-s} \sgn(\lambda_k). \ee
To be more precise, the ``limit'' here means the following: one first defines the sum on the right hand side for ${\mathrm{Re}}\, s$ sufficiently large that the series converges, and then (after proving
that this analytic continuation exists) one analytically continues to $s=0$.   

The derivation shows that for $m=-M$ and in the limit of large $M$,    ${\mathrm {Arg}}(Z_{\Psi,\APS})=-\pi \eta$, and since $|Z_{\Psi,\APS}|=1$, we get $\la\APS|Y\ra=Z_{\Psi,\APS}=\exp(-\i\pi\eta)$.
However, we have not yet defined $\eta$ precisely enough to take into account the fact that one or more of the $\lambda_k$ may vanish.  (Even if the four-dimensional Dirac operator $\D_W$ has
no zero-modes, which is the most relevant case for our derivation so far, the five-dimensional Dirac operator $\D_Y$ may have such modes.)   When we interpolate from $m=M$ to $m=-M$,
$\frac{-\i\lambda+m}{-\i\lambda+M}$ changes sign if $\lambda=0$, so a general formula for $\la\APS|Y\ra$ should include a factor of $-1$ for every zero-mode of $\i\sld_Y$.   We can achieve this
result
 by defining the function $\sgn(x)$ to include the case $x=0$:
\be\label{milko} \sgn(x)=\begin{cases} 1 & {\mathrm {if}}~x\geq 0 \cr
                                                              -1 & {\mathrm{if}}~x<0. \end{cases} \ee
 Then every zero-mode of $\sld_Y$ contributes $+1$ to $\eta$ and hence contributes a multiplicative factor $-1$ to $\exp(-\i\pi\eta)$.    We would get exactly the same result if we define $\sgn(0)=-1$;
    then a zero-mode of $\sld_Y$ contributes $-1$ to $\eta$ but still contributes                   a multiplicative factor $-1$ to $\exp(-\i\pi\eta)$.
    
    The conclusion then is a general formula for the partition function of the fermion field $\Psi$ on $Y$ with the boundary condition $\sL$ along $W=\partial Y$.   
    \be\label{nomme} Z(Y,\sL)=|\det\,\D^+_W|\exp(-\i\pi \eta).\ee
We can describe this by saying that ``anomaly inflow'' from $Y$, described  by the factor $\exp(-\i\pi\eta)$, is compensating for the anomaly of the chiral fermion that propagates along the boundary of $Y$.
This is a nonperturbative version of anomaly inflow; the inflow from the bulk  compensates for all perturbative and nonperturbative anomalies along the boundary.

The original formulation of anomaly inflow \cite{CH} involved perturbative anomalies only (triangle anomalies in the four-dimensional case) and used the Chern-Simons function instead of the eta-invariant.
Indeed, the APS formula for the change of $\eta$ under an infinitesimal change in the background fields shows that $\eta$ and Chern-Simons are equivalent perturbatively \cite{APS2}.   However,
for precise nonperturbative statements, and in particular to give a precise definition of the phase of the path integral of a chiral fermion, which we come to next,
it is important to use $\eta$, not Chern-Simons.              

As we explain next, for anomaly-free representations, the right hand side of eqn. (\ref{nomme})  actually does not depend on the choice of $Y$, and this expression
can be viewed as a formula for the chiral fermion partition function on $W$.                

\subsection{Anomalies and the Chiral Partition Function}\label{anomaly}

Now, under a certain topological restriction,
we can determine the condition for a chiral fermion system on $W$ to be anomaly-free, and, when it is anomaly-free, we can give a formula for the partition function of this system.
Specifically, we will  assume that the four-dimensional spin manifold $W$ and the gauge field $A$ on $W$ are such that $W$ is the boundary of a five-manifold  $Y$ and moreover the spin structure of $W$ and the gauge field $A$ both extend over $Y$.   In section \ref{goingon}, we will extend the analysis to cover the possibility that $Y$ may not exist, and we will see that
this does not lead to any new anomalies.   But for now, we assume that a suitable $Y$ exists.   If $Y$ does exist, there are many possible choices of $Y$.  For brevity, in the following
by ``the choice of $Y$,'' we will mean the choice of $Y$ together with the choice of an extension over $Y$ of the metric,  spin structure and gauge field of $W$.

We would like to interpret $Z(Y,\sL)$ as a formula for the chiral partition function on $W$.  A necessary condition for this is certainly that $Z(Y,\sL)$ must be independent of the choice of $Y$, since
in general there is not going to be any distinguished choice.   Let $Y$ and $Y'$ be two possible choices.   From eqn. (\ref{nomme}), we find that
\be\label{gifto} \frac{Z(Y,\sL)}{Z(Y',\sL)}= \exp(-\i \pi \eta_Y)\exp(\i \pi \eta_{Y'}), \ee
where $\eta_Y$, $\eta_{Y'}$ are the eta-invariants of $\D_Y$, $\D_{Y'}$, each with APS boundary conditions.
Now, let $\bar Y$ be the closed five-manifold that one can make by gluing together $Y$ and $Y'$ along their common boundary, after reversing the orientation of $Y'$.  The gluing
law for the eta-invariant \cite{DF} says that $\exp(-\i \pi \eta_Y)\exp(\i \pi \eta_{Y'})=\exp(-\i\pi\eta_{\bar Y})$, and therefore
\be\label{wifto}  \frac{Z(Y,\sL)}{Z(Y',\sL)}=\exp(-\i\pi\eta_{\bar Y}). \ee

In other words, a sufficient condition that $Z(Y,\sL)$ really depends only on $W$ and the fields on $W$, not on the choice of $Y$, is that for any closed five-manifold
$\bar Y$, with any spin structure and gauge field,  
we must have
\be\label{zolfo} \exp(-\i\pi \eta_{\bar Y}) = 1. \ee   This condition is also necessary; if there is a five-manifold $\bar Y$ that does not satisfy eqn. (\ref{zolfo}), then taking $W$ to be a small
sphere embedded in $\bar Y$ with $Y$ and $Y'$ the interior and exterior of the sphere, we get a case with $Z(Y,\sL)\not= Z(Y',\sL)$.   
Precisely when $\exp(-\i\pi \eta_{\bar Y})=1$ for all closed manifolds $\bar Y$, 
we can define the partition function of the fermion representation $R$ as $|\det\,\D_W^+|\exp(-\i\pi\eta)$, where this formula does not depend on the five-manifold
used to evaluate it.   

Let us see how the criterion (\ref{zolfo}) incorporates known anomalies, and makes it possible in many cases to show that there are no anomalies other than the known ones.
First of all, perturbatively $\eta$ is equivalent to Chern-Simons; more precisely, Atiyah, Patodi, and Singer \cite{APS2} showed that the change in $\eta$ under a small change in the metric or gauge field
is the same as the change in a corresponding Chern-Simons form, so locally in field space $\eta$ and Chern-Simons have the same dependence on the metric and gauge field and they capture the
same perturbative anomalies. 
The perturbative anomaly of any chiral fermion representation  $R$ in four dimensions is related via anomaly inflow to a certain Chern-Simons form in five dimensions.
Precisely when $R$ has no perturbative anomaly, the Chern-Simons form vanishes.   
  In this case, $\exp(-\i\pi \eta_{\bar Y})$ is invariant under continuous variations of the metric and gauge field,
and thus is a topological invariant.\footnote{Although $\exp(-\i\pi\eta_{\bar Y}) $ is a topological invariant, $\eta_{\bar Y}$ is not: it jumps by 2 when an eigenvalue of $\i\sld_Y$ changes sign.}

Let us now see how $\exp(-\i\pi\eta_{\bar Y})$ detects a global or nonperturbative anomaly in a case in which there is one.   We take the gauge group to be $G=SU(2)$ (a symplectic group
$Sp(k)$ would be similar; note that $SU(2)\cong Sp(1)$).  As $G$ does not
have complex representations, in four dimensions all representations of $G$ are free of perturbative anomalies.   Now let us evaluate $\exp(-\i\pi\eta_{\bar Y})$ for the case that the representation
$R$ is a representation of half-integer spin, for example the spin 1/2 representation.   The one fact that we will use about such representations is that they are pseudoreal.\footnote{$SU(2)$ representations
of integer spin are real. A real representation
of $SU(2)$ -- or indeed of any group -- allows a bare mass, so it is completely anomaly-free.}

The dimension four spinor representation $S$ of ${\mathrm{Spin}}(5)$ is likewise pseudoreal.
  The Dirac operator $\sld_Y$ valued in $R$ then acts on a wavefunction that is valued in the tensor product $S\otimes R$, and, as the tensor product of two pseudoreal representations,
  this representation is real.
This means that $\sld_Y$ can be defined as a real operator.  So the self-adjoint Dirac operator $\i\sld_Y$ is imaginary and anticommutes with complex conjugation, which we will denote
as $\star$.   If  $\i\sld_Y \Psi=\lambda \Psi,$ then $\i\sld_Y(\star \Psi)=-\lambda \star\Psi$.   Thus the nonzero eigenvalues of $\i\sld_Y$ come in pairs $(\lambda,-\lambda)$, and since
such a pair makes no net contribution to $\eta_{\bar Y}$, it follows that $\eta_{\bar Y}$ can be evaluated just from the contributions of zero-modes of $\sld_Y$.   Let $\zeta$ be the number of such
zero-modes.   Each one contributes $+1$ to $\eta$, so in this situation
\be\label{polgo}\exp(-\i\pi\eta_{\bar Y})= (-1)^\zeta. \ee

Indeed, it was found in \cite{Witten1982} that a representation of $SU(2)$ is anomalous if the five-dimensional Dirac operator can have an odd number of zero-modes.
The original example was a certain  $SU(2)$ gauge field over $S^5$ with $\zeta=1$ for the spin 1/2 representation and more generally for representations of spin $2n+1/2$ (but not $2n-1/2$).      Another simple example, discussed in \cite{NewSU2}, is the following.
Take $\bar Y=S^4\times S^1$ with a product metric, and a spin structure that is periodic in the $S^1$ direction; assume that the gauge field $A$ has instanton number 1 on $S^4$ and is pulled back from $S^4$
(thus $A$ is invariant under rotation of $S^1$ and has no component in the $S^1$ direction).   Then a zero-mode of the Dirac operator on $S^4\times S^1$ has no dependence on $S^1$ and is derived from
a zero-mode of the Dirac operator on $S^4$.   So there is an anomaly if (counting both chiralities of fermion) 
the Dirac operator on $S^4$, in a field of instanton number 1, has an odd number of zero-modes.

Are these two examples equivalent?   Do they detect anomalies for the same representations of $SU(2)$?    This
question, of course, can be decided by computation.   What cannot be so easily decided by computation is the following question: are there other examples that are independent of these two?
To show that a given representation is completely
anomaly-free, must we consider five-manifolds other than  $S^5$ and $S^4\times S^1$?

To answer this type of question, it is useful to know that
 $\exp(-\i\pi\eta_{\bar Y})$ is not just a topological invariant, but  a cobordism invariant.   This means that $\exp(-\i\pi\eta_{\bar Y})=1$ if $\bar Y$ is the boundary of a six-manifold
$Z$ over which the spin structure and gauge field extend.   That claim can be proved by considering the  index theorem for the index $\I$ of the Dirac operator $\sld_Z$ on $Z$, again
with values in the representation $R$.   In general, the  Atiyah-Singer index theorem for the Dirac operator on a closed manifold $Z$ of any even dimension $2r$  has the general form $\I=\int_Z \Phi$, where $\Phi$ is a differential form that is a polynomial of degree $r$ in the Riemann tensor of $Z$ and the Yang-Mills field strength.   If, however, $Z$ has
boundary $\bar Y$, then the formula receives a boundary contribution and becomes
\be\label{inbo} \I=\int_Z\Phi-\tfrac{1}{2}\eta_{\bar Y}. \ee
This is known as the APS index theorem.  In our application,  $\Phi$ represents the six-dimensional characteristic class that is related to the anomaly Chern-Simons 
five-form  in five dimensions.  So if we started with an anomaly-free
representation $R$ in four dimensions, then $\Phi=0$ and the APS index theorem reduces to ${\mathcal I}=-\tfrac{1}{2}\eta_{\bar Y}$.   As $\mathcal I$ is an integer, it follows that $\eta_{\bar Y}$ is
an even integer and $\exp(-\i\pi\eta_{\bar Y})=1$.  

Manifolds of some given dimension with some additional structures such as a spin structure, a gauge field, etc., can  be made into a group. The group
law is disjoint union of manifolds, and one imposes an equivalence relation that a manifold is considered trivial if it is the boundary of a manifold of one dimension more
over which the structures extend.   The group defined this way  is called the cobordism group.   The fact that $\exp(-\i\pi\eta_{\bar Y})=1$ if $\bar Y$ is the boundary of some $Z$ (over which all structures extend)
means that the map  $\bar Y\to \exp(-\i\pi\eta_{\bar Y})$ can be viewed as a group homomorphism from the cobordism group to $U(1)$.   Since cobordism groups often have been or can be determined
mathematically, this often gives a practical way to identify the possible anomalies in a given example.  For instance, in \cite{WittenOtherOld}, this approach was used to show that the low energy supergravity
associated to the heterotic string is completely anomaly-free; the relevant cobordism group was shown to vanish in an appendix to the paper by R. E. Stong.  

Going back to $SU(2)$ gauge theory, the cobordism group that classifies five-dimensional spin manifolds with an $SU(2)$ bundle is $\Z_2$, generated by either of the two examples discussed earlier.
So those examples are equivalent to each other, and there are no other examples to check.

For more examples of such arguments in dimensions 1, 2, 3, and 4, see section 4 of \cite{WY}.   In particular, it is shown there that in four-dimensional gauge theory with
a connected and simply-connected gauge group, there are no anomalies except the usual triangle anomaly, and the mod 2 anomaly that arises when there can be an odd
number of fermion zero-modes in an instanton field.  Among simple Lie groups, that latter anomaly arises only for symplectic groups $Sp(k)$, and in particular, as just discussed, for 
$SU(2)\cong Sp(1)$.

 \subsection{Theta-Angles And Fermion Zero-Modes}\label{goingon}
 
 There are two deficiencies in the explanation that we have given so far.   First, we assumed that the four-manifold $W$ on which we wish to study a given chiral gauge theory
 is the boundary of a five-manifold $Y$ over which the spin structure and gauge field of $W$ extend.   In other words, we assumed that $W$ is trivial as an element of the relevant
 cobordism group.    Second, we assumed that the partition function of the chiral fermions on $W$ is generically nonzero.  
 
  The importance of this second assumption is that if the partition
 function is generically nonzero, then the path integral of fermions coupled to  background gauge and gravitational fields is adequately described by giving the formula (\ref{nomme}) for
 that partition function.    The only additional information that one needs to compute a path integral with insertions of operators containing fermions is a knowledge of the fermion
 propagator, and in the absence of zero-modes, this is always well-defined for any representation, anomaly-free or not.   However, if the fermion partition function  vanishes identically,
 then to characterize the fermion path integral, we need to replace eqn. (\ref{nomme}) by a similar formula for a correlation function that is generically nonzero.  Such a formula will be described
 at the end of this section.
 
 The two obstructions are closed related, since when the fermion partition function  vanishes identically due to an index theorem (or, in some spacetime dimensions, a mod 2 index
 theorem), this always represents an obstruction to null-cobordism of $W$, that is, to the existence of $Y$.   However, the cobordism obstruction is more general since it is possible
 to have obstructions in the cobordism problem that are not reflected in fermion zero-modes.   For example, in four dimensions, this can happen in chiral gauge theories
with suitable gauge groups.   Therefore, we will consider first the cobordism problem, ignoring the second question about fermion zero-modes, and then explain how
 to modify the definitions to take into account the fact that the fermion path integral vanishes identically in some topological sectors.
 
In four dimensions, the obvious obstructions in the cobordism problem are given by gauge and gravitational instantons.  Suppose, for example, that  $F$ is
the gauge field strength in a simple factor of the gauge group.   Then if $W$ is the boundary of some $Y$ over which the gauge field extends, we have $\int_W \Tr\, F\wedge F=
\int_Y \d\Tr\,F\wedge F=0$, where we use Stokes's theorem and the Bianchi identity.   So a nonzero gauge theory instanton number is an obstruction to existence of $Y$.
Replacing $F$ with the Riemann tensor $R$, the same reasoning applies to show that a nonzero $\int_W\Tr\, R\wedge R$, representing  the gravitational instanton number,
 is also an obstruction to existence of $Y$.  In the group
theory sense, these
obstructions are of infinite order, since if $W$ has a nonzero gauge or gravitational instanton number, the disjoint union of any (nonzero)  number of copies of $W$ has the same obstruction and is still  not the boundary of
any $Y$.   When we say ``any number of copies,'' we include the possibility of a negative number of copies.  Here we define $-W$ to be $W$ with orientation reversed (but no change
in gauge field or spin structure), and by $-n$ copies of $W$, we mean $n$ copies of $-W$. Note that $-W$ is always the inverse of $W$ in the cobordism group, since the disjoint union $W+(-W)$ 
is the boundary of $W\times I$, where $I$ is an interval.   (After picking an orientation of $W\times I$ and using a definite recipe like ``outward normal first'' to orient
   the boundary of an oriented manifold, one finds that the two boundaries of $W\times I$ are oppositely oriented.)  Indeed, it is this fact that turns the cobordism relation into a group.
   
     In four dimensions, gauge and gravitational instantons are the only obstructions of infinite order in the cobordism
problem.   In four-dimensional gauge theories with connected and simply-connected gauge groups, there are no obstructions of finite order, but that is not so for other gauge groups.
 For example, in $SU(N)/\Z_N\times SU(N)/\Z_N$ gauge theory in four dimensions, there is an obstruction
of order $N$ which  can be described as follows.   An $SU(N)/\Z_N$ bundle $E\to W$ has a mod $N$ characteristic class $\upalpha\in H^2(W,\Z_N)$ ($\upalpha$ is the second Stieffel-Whitney class of $E$ if $N=2$), so in $SU(N)/\Z_N\times SU(N)/\Z_N$ gauge
theory, there are two such classes $\upalpha$ and $\upalpha'$, and there is a mod $N$ invariant $\int_W\upalpha\cup\upalpha'$ that does not contribute to any index formula
and is not reflected in any fermion zero-modes.   In four dimensions, other examples can be given with finite gauge groups or more general gauge groups that are not connected.

In general, the cobordism group that classifies obstructions to the existence of $Y$ is a finitely generated abelian group.   Such a group has the general form
$\Z^k\times \Z_{r_1}\times \Z_{r_2}\times\cdots\times \Z_{r_s}$, for some $k$ and $r_1,\cdots, r_s$.   In such a case, we can pick generators $W_1,W_2,\cdots,W_{k+s}$
for the cobordism group; here $W_1,\cdots, W_k$ are of infinite order and for $i>0$, $W_{k+i}$ is of order $r_i$.
   Each of these generators is a four-manifold with some spin structure and gauge field.  The statement that they generate the cobordism group means that any four-manifold
   $W$ with spin structure and gauge field is cobordant to a disjoint union $\oplus_{i=1}^{k+s} n_i W_i$, with integers $n_i$ that are uniquely determined if $i\leq k$
   and uniquely determined mod $r_j$ if $i=k+j$, $j>0$.
   
   None of the generators $W_i$ are boundaries of any $Y$, so a first task we might consider is how to define the chiral fermion partition function on $W_i$ for an anomaly-free representation.
   First let us consider  $W_{k+i}$, which  is of some finite order $r_i$.   Then the disjoint union of $r_i$ copies of $W_{k+i}$ is the boundary of some $Y$.   So we can use the formula
   (\ref{nomme}) to determine the partition function of chiral fermions on the disjoint union of $r_i$ copies of $W_{k+i}$.  Since physically, the partition function of a quantum
   field theory on the disjoint union of manifolds is the product of the individual partition functions, we interpret this partition function for $r_i$ copies of $W_{k+i}$ as the  $r_i^{th}$ power of the
   partition function $Z(W_{k+i})$ for a single copy.    Given this formula for $Z(W_{k+i})^{r_i}$, how do we determine $Z(W_{k+i})$?   Of course, we want to take an $r_i^{th}$ root.  There
   are $r_i$ possible choices of $r_i^{th}$ root, and each choice will lead to a sensible definition of the chiral fermion partition function.   In fact, if there is an obstruction of order $r_i$ in the cobordism problem for some
   gauge theory, then the gauge theory will have a $\Z_{r_i}$-valued discrete theta-angle.   The procedure just described gives a precise explanation of how to incorporate
this theta-angle at the quantum level.   The subtlety in defining the phase of a chiral fermion partition function will generally frustrate more obvious attempts to do that.
   
   Now let us consider one of the $W_i$ that is of infinite order in the cobordism group.   This means that no multiple of $W_i$ is a boundary.   However, as previously noted, $W_i+(-W_i)$
   is the boundary of $W_i\times I$.    So we can use eqn. (\ref{nomme}) to determine the chiral fermion partition function
   on the disjoint union $W_i+(-W_i)$.   As before, since the path integral  on a disjoint union is the product of path integrals, we interpret the resulting formula for the path integral\footnote{In four dimensions,
   the generators of the cobordism group of infinite order are  related to the gauge and gravitational instanton numbers.  This means that typically, the chiral fermion partition functions on the infinite
   order generators
   $W_i$ are identically zero.   We ignore this momentarily and  explain presently how to modify the procedure to take this into account.  In four dimensions, there is no zero-mode problem for the generators of finite order, because
   there is no mod 2 index theorem.}
   $Z(W_i+(-W_i))$ 
   of the disjoint union $W_i+(-W_i)$ as a formula for the product $Z(W_i)Z(-W_i)$.  A  general principle of Euclidean quantum field theory, related to reflection positivity, is that reversing
   the orientation of spacetime with no change in the other data (in the present context, the metric, spin connection, and gauge field on $W$) has the effect of complex conjugating the partition function.
   So we interpret $Z(W_i)Z(-W_i)$ as $|Z(W_i)|^2$.   Thus, we get a natural definition of the absolute value $|Z(W_i)|$.   We do not get any natural definition of the phase of $Z(W_i)$, so we just
   assign it an arbitrary phase.    So we learn that to  every generator of infinite order in the cobordism group, there is a corresponding continuous theta-angle.   In the case of chiral
   gauge theories in four dimensions, the generators of the cobordism group of infinite order are associated to the gauge theory and gravitational instanton numbers.   For each instanton type,
   there is an associated theta-angle.  These theta-angles can be incorporated  at the quantum level in the manner just described.   The subtlety of defining the phase of the chiral
   fermion path integral will generally frustrate more obvious attempts to define the renormalized theta-angles, provided that one wants to be able to define the theory on a general four-manifold.
   However, if one wishes to work only on $T^4$, then for connected, simply-connected gauge groups,
   one can use cluster decomposition instead, as in \cite{Craig}.\footnote{For discrete theta-angles, cluster decomposition does not appear to provide
   a satisfactory shortcut to the argument based on cobordism, even on $T^4$. That is roughly because the gauge field configurations that carry discrete topological obstructions like $\int_W\upalpha\cup\upalpha'$
   cannot be localized in the same way that instantons can be localized.  There actually is a problem even in the absence of discrete theta-angles if fractional instanton number on $T^4$ is possible,
   as for gauge group $SU(N)/\Z_N$, because the fractional part of the instanton number cannot be localized and treated by cluster decomposition.}
   
   Now we are ready to explain how to define the chiral fermion path integral for any $W$.   By definition of the generators, $W$ is cobordant to some disjoint union $\sum_i n_i W_i$,
   which means that a disjoint union $W-\sum_i n_i W_i$ is the boundary of some $Y$.  (For $i\leq k$, $n_i$ is uniquely determined, but $n_{k+i}$ is uniquely determined mod only $r_i$.) So eqn. (\ref{nomme}), or more exactly its extension described shortly to accomodate fermion zero-modes,
   gives a definition of the chiral fermion path integral on $W-\sum_i n_i W_i$, which we interpret as $Z(W) \prod_i Z(-W_i)^{n_i}=Z(W) \prod_i \overline{Z(W_i)}^{n_i}$.  Since the $Z(W_i)$ have already
   been defined, this gives
   a general definition of the chiral fermion path integral for any $W$, in terms of the quantum renormalized theta-angles.   As usual, this definition does not depend on the choice of $Y$, and using the fact that $r_i$ copies of $W_{k+i}$ are a boundary, one can verify that the definition in unaffected if $n_{k+i}$ is shifted by a multiple of $r_i$.
A similar argument  was used in \cite{WittenDInst} to define precisely the worldsheet path integral of the heterotic string in a general
   background.

   Part of what we have learned is that  the discrete and continuous theta-angles, taken together, are associated  to homomorphisms from the relevant four-dimensional
   cobordism group to $U(1)$.  A precise statement is that given any such homomorphism $\Upsilon$, the theta-angles can be shifted by multiplying the partition function $Z(W)$ by an additional
   factor $\Upsilon(W)$.      
   
   Now we will consider the case that the chiral fermion partition function vanishes identically on $W$ because of zero-modes predicted by an index theorem (or, in some dimensions, a mod 2 index theorem).
   When this happens, $W$ is not a boundary, so we have to apply the machinery with a bounding five-manifold to a disjoint union of $W$ with other components, for example $\bar W=W-\sum_i n_i W_i$,
   as just discussed.    We cannot now characterize the fermion path integral on $\bar W$ by the formula (\ref{nomme})  for the partition function, since the partition function vanishes identically.  Instead we need a formula
   generalizing that equation but describes path integrals on $\bar W$ with operator insertions chosen to get a non-zero result.   We will describe how to construct a relatively direct analog of the previous formula.   For an alternative explanation in a more general context, see  section 2.5 of \cite{WY}.
   
   First of all, the definition of APS boundary conditions that we gave previously assumed that the operator $\D_W$ has no zero-modes.   Under this condition, APS boundary conditions are described
   by saying that the coefficients of eigenmodes of $\D_W$ with negative eigenvalue are required to vanish along the boundary.     To generalize APS boundary conditions without this restriction,
   we need to place a constraint on the coefficients of the zero-modes.    We want a condition that will make $\D_Y$ self-adjoint, so that it will have a discrete spectrum of real eigenvalues, just
   as when $\D_W$ has  no zero-modes.   It turns out that a self-adjoint boundary condition of this type exists if and only if the index of the chiral Dirac operator $\D_W^+$ vanishes.   We are now
   in a situation in which this is not true on the connected manifold $W$, but it is true on the disconnected manifold $\bar W=W-\sum_i n_i W_i$.   So, if $\bar W=\partial Y$, we will have to complete the definition 
   of APS boundary conditions by defining a boundary condition along $\bar W$ that mixes zero-modes on different boundaries.   The necessary procedure is as follows.
   Since the index of $\D_{\bar W}^+$ vanishes, the spaces $V_+$ and $V_-$ of positive and negative chirality zero-modes of $\D_{\bar W}$ have equal dimension.   So there exist unitary
   isomorphisms $U:V_+\cong V_-$.   For any such choice, we complete the definition of the APS boundary condition on $\i\sld_Y$  by saying that the coefficients of positive and negative chirality zero-modes of
   $\D_W$ are related by\footnote{We can motivate this definition by observing that near $\bar W$, $\i\sld_Y$, restricted to the zero-modes of $\D_{\bar W}$,
   can be approximated by a 1-dimensional Dirac operator  $\i\sld_Y= \i \begin{pmatrix}\frac{\d }{\d \tau} & 0\cr 0 & -\frac{\d}{\d \tau}\end{pmatrix}\otimes I_{\N}$, where $I_{\N}$ is the identity on an $\N$-dimensional
   vector space; here upper and lower components correspond to positive and negative chirality zero-modes of $\D_{\bar W}$, and we have incorporated the fact that these are equal in number as
   $\D_{\bar W}^+$ has zero index. (We group all zero-modes of $\D_{\bar W}$ together here, regardless of which boundary they come from.)   It is fairly well known that the general self-adjoint boundary condition on this operator  restricts the wavefunction $\Psi=\begin{pmatrix}\psi_+\cr \psi_-\end{pmatrix}$
   so that $\left.\psi_-\right|_{\tau=0}=U \left.\psi_+\right|_{\tau=0}$ for some unitary matrix $U$, and this is the APS boundary condition.} $U$.   This gives a self-adjoint boundary condition that has all essential properties that we used in section \ref{startup}.
   In particular, the boundary condition determines a state $|\APS\ra$ in the boundary Hilbert space $\H_{\bar W}$, which has a well-defined phase, since, if $\bar W=\partial Y$,
    the inner product $\la\APS|Y\ra$ with the state
   $|Y\ra$ created by the path integral on $Y$ is simply given by a path integral on $Y$ with APS boundary conditions.   In fact, with mass parameter $m\cong -M$ and  cutoff $M$ taken to infinity,
   $\la\APS|Y\ra$ is given by the same formula as before, with the same derivation:
   \be\label{jubo} \la\APS|Y\ra =\exp(-\i\pi \eta_Y). \ee
   The eta-invariant, of course, is now computed using the eigenvalues of $\D_Y$ with the generalized APS boundary conditions.
   
   There is no problem to define the local boundary condition $\sL$ for $W$ and for all of the $W_i$.   However, in trying to repeat the derivation of eqn. (\ref{bolme}), we run into the fact that $\la\sL|\Omega\ra=0$,
   where $\Omega$ is the fermion ground state.   This is related, of course, to the fact that the partition function $Z(Y,\sL)$ on the left hand side of eqn. (\ref{bolme}) also vanishes.   To characterize the
   fermion path integral, we need a formula not for the partition function but for the unnormalized expectation value of some operator $O$ that has a generically nonzero expectation value.  Such an operator will have to 
   ``absorb'' the fermion zero-modes, and a minimal choice of $O$  on any given manifold is the product of a minimum number of elementary  fermion fields that generically suffices to absorb zero-modes on that manifold.
   (This number can be determined by applying the index theorem to irreducible components of the representation $R$.)    If a manifold $\bar W$ is the disjoint union of components $W_\alpha$ (for example, $W$ and some of the $W_i$ or $-W_i$), we can
   define a minimal observable $O_\alpha$ to generically absorb zero-modes on $W_\alpha$, and then we set $O=\prod_\alpha O_\alpha$.   If the fermion representation is anomaly-free, there is
    no factor-ordering issue here;
   each $O_\alpha$ is bosonic because in any anomaly-free representation, the number of fermion zero-modes on a manifold is always even (this may seem clear from the discussion of $SU(2)$ gauge theory in section \ref{anomaly}, and is proved at the end of this section).    For such $O$, the path integral $\la O\ra_{Y,\sL}$ on some $Y$ of boundary $\bar W$, with boundary condition $\sL$ and boundary insertion $O$, is generically nonzero, and
we want a formula for it generalizing eqn. (\ref{nomme}).

We can think of the  fermion fields that appear in the operators $O_\alpha$
as the massless boundary fields $\chi$ and $\bar\chi$ that were introduced in eqns. (\ref{poly}) and (\ref{noly}); of course, for $|m|,M<\infty$, these are modes of
the five-dimensional fields $\Psi,\bar\Psi$.    
Suppose that on $W_\alpha$, $\chi$ and $\bar\chi$ generically have $\mu_\alpha$ and $\nu_\alpha$ zero-modes respectively.   As we prove later, $n_\alpha=\mu_\alpha+\nu_\alpha$ is
always even in the case of an anomaly-free representation, but separately $\mu_\alpha$ and $\nu_\alpha$ can be odd.\footnote{It is possible for $\mu_\alpha$ and $\nu_\alpha$ to be both nonzero
and in fact odd
because the representation $R$ may be reducible, and the gauge group $G$ is not necessarily simple; there can be multiple contributions to the index theorem, from  gravitational instantons and
different types of gauge instanton.
For a simple example, let $G=SU(N)\times SU(N)\times SU(N)$, and take $R$ to be the chiral but anomaly-free representation $(N,\bar N, \1)\oplus (\1,N,\bar N)\oplus (\bar N,\1,N)$.  Let $\bar W$ be
a disjoint union $W_1+W_2+W_3$, with each of the $W_i$ a copy of $T^4$, and take the instanton numbers to be $(1,-1,0)$ on $W_1$, $(-1,0,0)$ on $W_2$, and $(0,1,0)$ on $W_3$, a choice
that is consistent with $\bar W=\partial Y$.   Then $(\mu_1,\nu_1)=(N,N)$, $(\mu_2,\nu_2)=(0,2N)$, $(\mu_3,\nu_3)=(2N,0)$. So $\mu_i+\nu_i$ is even for all $i$, but $\mu_1$ and $\nu_1$ are odd if $N$ is odd.
And $\sum_i\mu_i=\sum_i\nu_i$.}   Since the index of $\D_{\bar W}^+$ is assumed to vanish, we can define $\N=\sum_\alpha \mu_\alpha
=\sum_\alpha \nu_\alpha$.

Let $\psi_i$, $i=1,\cdots,\N$, be an orthonormal basis of the positive chirality zero-modes of $\D_{\bar W}$, corresponding to zero-modes of $\chi$,
and set $\psi'_i=U\psi_i$, with $U$ the unitary map that enters the APS
boundary condition.  The $\psi'_i$ are an orthonormal basis of the negative chirality zero-modes of $\D_{\bar W}$ in the representation $R$, and their complex conjugates $\bar\psi'_i$
are the zero-modes of $\bar\chi$.  (It is natural, though not necessary, to take each $\psi_i$ to be supported on a single component $W_\alpha$; the $\psi'_i$ then generically do not have that property.)  We expand
\be\label{expand}\chi=\sum_i \chi_i\psi_i+\cdots,~~~~\bar\chi=\sum_i\bar\chi'_i\bar\psi'_i+\cdots,\ee with anticommuting  coefficients $\chi_i,\bar\chi_i'$,
where the omitted terms are nonzero-modes.
Up to a sign $(-1)^\Delta$ that comes from reordering the factors in $O=\prod_\alpha O_\alpha$, we can write
$O=\chi(x_1)\cdots\chi(x_\N)\bar\chi(y_1)\cdots\bar\chi(y_\N)$, with points $x_j,y_j\in\bar W$ (the correct number of them in each $W_\alpha$).
As $O$ contains exactly as many fermion fields as there are zero-modes, only the zero-mode parts of these fields contribute to $\la O\ra_{Y,\sL}$, and inside the path
integral we can replace $O$ by
\be\label{replo} (-1)^\Delta\det[\psi_i(x_j)]\det[\bar\psi'_i(y_j)]\,\chi_1\cdots\chi_\N\bar\chi'_1\cdots\bar\chi'_\N. \ee
The determinants come from summing over the ways that $\N$ elementary fermion fields can absorb $\N$ zero-modes.
So it suffices to compute the path integral with insertion of $\chi_1\cdots\chi_\N\bar\chi'_1\cdots\bar\chi'_\N$.

For this, we perturb the operator $\D_{\bar W}$ to lift its zero-modes.
  Let $\Pi_\pm$ be projection operators from the spaces of positive
or negative chirality spinors to the zero-mode subspaces.   For small $\epsilon>0$, define the self-adjoint operator $\D_{\bar W,\epsilon}=\D_{\bar W}+\epsilon (U\Pi_++U^\dagger \Pi_-)$.  This operator
has no zero-modes, since for instance it maps a positive chirality zero-mode $\psi$ to the negative chirality zero-mode $\epsilon U\psi$. 
By extending $U$ slightly into $Y$, we can define a self-adjoint operator $\D_{Y,\epsilon}$ on $Y$ that restricts to $\D_{\bar W,\epsilon}$ on $\bar W$.
 On $\bar W$, if  $\Psi=\begin{pmatrix}\psi\cr\psi'\end{pmatrix}$, where $\psi$
and $\psi'$ are zero-modes of $\D_{\bar W}$ with positive and negative chirality, then $\Psi$ is a linear combination of eigenstates of $\D_{\bar W,\epsilon}$ of eigenvalue $\pm\epsilon$, and the condition
that $\Psi$ has eigenvalue $+\epsilon$ is that $\psi'=U\psi$.      So APS boundary conditions for $\D_{Y,\epsilon}$  coincide with the APS boundary conditions that we described earlier at $\epsilon=0$  for
the original $\D_Y$, with the same $U$.  As $\D_{\bar W,\epsilon}$ has no zero-modes, we can simply use eqn. (\ref{nomme}) to compute $Z(Y,\sL)_\epsilon$.  Since $\D_{\bar W,\epsilon}^{+\dagger }\D_{\bar W,\epsilon}^+$ has $\N$ eigenvalues
$\epsilon^2$ and otherwise has the same spectrum as $\D_{\bar W}^{+\dagger}\D_{\bar W}^+$, we have $|\det\,\D_{\bar W,\epsilon}^+|=\epsilon^\N |\det'\, \D_{\bar W}^+|$, where $\det'$ is a regularized determinant
computed in the space orthogonal to the zero-modes.   So to leading order in $\epsilon$,
\be\label{friendly} Z(Y,\sL)_{\epsilon}= \epsilon^\N\prod_\alpha |{\mathrm{det}}'\D_{W_\alpha}^+| \exp(-\i\pi \eta_Y), \ee
where $\eta_Y$ is the eta-invariant of the original operator $\D_{Y}$ at $\epsilon=0$, with APS boundary conditions defined by $U$.  On the other hand, the perturbation has added to the action
a term $-{\epsilon}\sum_i\bar\chi'_i\chi_i$, so the coefficient of $\epsilon^\N$ in  $Z(Y,\sL)_\epsilon$ is
the path integral on $Y$ at $\epsilon=0$ with insertion of $\prod_i \bar\chi'_i \chi_i=(-1)^{\N(\N+1)/2}\chi_1\cdots\chi_\N\bar\chi'_1\cdots\bar\chi'_\N$. Combining this with (\ref{replo}), we get finally\footnote{A careful
reader may notice that we have elided a detail.   If normalized four-dimensional zero-modes are related to five-dimensional zero-modes via eqns. (\ref{poly}) and (\ref{noly}), then the five-dimensional
modes have norm squared 
$\frac{1}{2|m|}$.   As a result, in the derivation  just sketched, each zero-mode should come with a factor $\frac{\epsilon}{2|m|}$ (not just $\epsilon$).   
As shown in \cite{WY} and remarked in footnote \ref{lookback},
each nonzero mode also comes with a factor $\frac{1}{2|m|}$.  The product of $\frac{1}{2|m|}$ over all modes disappears after regularization and renormalization.   If we did not take into account the
factors coming from zero-modes, we would be left with an uncompensated factor $(2|m|)^{\N}$.}
 \be\label{xinco} \la O\ra_{Y,\sL}   =(-1)^{\Delta+\N(\N+1)/2} \prod_\alpha  |{\mathrm{det}}' \,\D_{W_\alpha}^+|\exp(-\i\pi \eta_Y)\det[\psi_i(x_j)]\det[\bar\psi'_i(y_j)].\ee

Two consistency checks are worth noting.  First, the right hand side of (\ref{xinco}) does not depend on the choice of basis $\psi_i$.  Indeed,
if we transform the $\psi_i$ by $\psi_i\to \sum_j \psi_j V_{ji}$ with unitary $V$, then  $\psi'_i\to\sum_j\psi'_j V_{ji}$, so the first determinant is multiplied by $\det\,V$ and the second by $\det\,\bar V=(\det\,V)^{-1}$.
Second, the left hand side of (\ref{xinco}) is defined without any reference to APS boundary conditions, so it does not depend on $U$.  On the right hand side, $U$ enters both in the
eta-invariant and through $\psi'_i=U\psi_i$, and as $U$ varies, $\det[\bar\psi'_i(y_j)]$ varies as a constant times $(\det\,U)^{-1}$.  So $\exp(-\i\pi\eta_Y)$ must vary as a
constant times $\det\,U$.  This is part of the Dai-Freed theorem \cite{DF}.

Assuming that the representation $R$ is anomaly-free, by the usual argument the right hand side of the formula (\ref{xinco}) does not depend on $Y$. So the formula gives us information about the purely four-dimensional theory on $\bar W= \partial Y$.
 As before, the only problem is that $\bar W$ is the union of disjoint components $W_\alpha$, and we want to define the chiral gauge theory on a single component.
 For this we specialize to the case
  $\bar W=W-\sum_i n_i W_i$,
 where $W$ is the manifold of interest and $W_i$ are  generators of the cobordism group.   For each $W_i$, we pick as before a fiducial metric and gauge field on each $W_i$, but now
 we also pick a fiducial observable $O_i$ of the previously described type on each\footnote{If $W_i$ is not a torsion generator, let $O_i$ and $O_i^-$ be fiducial observables on $W_i$ and on $-W_i$.
 We want to ensure that the product $\la O_i^-\ra_{-W_i}
 \la O_i\ra_{W_i}$, computed by applying eqn. (\ref{xinco}) on $Y=W_i\times I$, is positive, so that we can take a square root to define $|\la O_i\ra_{W_i}|$.  Positivity will follow from reflection positivity applied to the reflection 
 that exchanges the two ends of $Y$ if the fiducial observables and the APS boundary conditions are chosen to respect reflection positivity.
 For this aim, having picked a fiducial observable $O_i$ on $W_i$, we pick on $-W_i$ the fiducial observable $O_i^-$ that is obtained from $O_i$ by the operation corresponding
 to reflection positivity (complex conjugation and reversal of orientation).   
We also define the APS boundary condition by taking $U$ to be the identity map from zero-modes on $W_i$ to zero-modes on $-W_i$. }   $W_i$.     Then reasoning as before, the formula (\ref{xinco}) determines 
 $\la O_i\ra_{W_i}$ up to a phase if $i\leq k$, and determines $\la O_{k+i}\ra_{W_{k+i}}$ up to an $r_i^{th} $ root of unity.   Fixing the undetermined constants amounts to specifying the values
 of renormalized theta-angles -- both discrete ones and continuous ones.   Once all the theta-angles are fixed, eqn. (\ref{xinco}) contains enough information to define
 $\la O\ra_W$ on each four-manifold $W$.   And once this is determined, arbitrary chiral fermion correlation functions on $W$ are uniquely determined by using
 this information plus the fermion propagator defined in a space orthogonal to the zero-modes; that propagator is always well-defined for any representation, anomaly-free or not.

  To avoid a troublesome minus sign in defining $\prod_\alpha O_\alpha$, we claimed that in any anomaly-free representation, the number $\zeta$ of fermion zero-modes on any  manifold $W$ is always even.   This may be proved as follows.
First, let $Y=W\times S^1$ with a product metric.  Parametrize $S^1$ by a periodic variable $\tau$ with $\tau\cong \tau+L$, and pick a spin structure in which fermions are periodic under $\tau\to\tau+L$. Let $x$
   denote local coordinates on $W$.  Assume the gauge field on $Y$ is pulled back from $W$.    The five-dimensional Dirac operator $\D_Y$ is odd under the reflection symmetry 
   $\Psi(x,\tau)\to \gamma_\tau \Psi(x,-\tau)$.   So nonzero eigenvalues come in pairs $(\lambda,-\lambda)$, and therefore $\eta(\D_Y)$ is simply the number of zero-modes of $\D_Y$.
   A zero-mode of $\D_Y$  must have zero momentum in the $S^1$ direction, so it must come from a zero mode of $\D_W$.   Hence $\eta(\D_Y)=\zeta$ and $\exp(-\i\pi\eta(\D_Y))=(-1)^\zeta$.
   So $\zeta$ must be even to avoid an anomaly.   This argument is obviously reminiscent of the  discussion of $SU(2)$ gauge theory in section \ref{anomaly}, but in that case, the proof that nonzero
   modes of $\D_Y$ are in pairs $(\lambda,-\lambda)$ required an assumption that the fermion representation is pseudoreal with no restriction on $Y$, while 
   here there is no restriction on the fermion representation but we require a specific form of $Y$.

\section{Continuum Interpolations of Admissible Lattice Gauge Fields}\label{interpolation}

As noted in section \ref{facttwo}, continuum interpolations of admissible nonabelian lattice gauge fields have been constructed in several papers, including  \cite{GKSW,PS2,Hooft,HS} and recently \cite{Craig}, in varying degrees of generality.
   There are various routes to construct such an interpolation for arbitrary compact $G$ (not necessarily connected or  simply-connected).   One route -- an extension of that  of 't Hooft \cite{Hooft} -- will be briefly described here. Though we phrase the following initially for the four-dimensional case, it will be evident that the 
arguments apply equally well in any dimension $d$.  In fact, as we will explain, the procedure can be carried out on any $d$-manifold.

For orientation, we can start with a hypercubic lattice in $T^4$.    The lattice divides $T^4$ into four-dimensional hypercubes.   For $0\leq k\leq 3$, the $k$-dimensional faces of such a hypercube
are themselves $k$-dimensional hypercubes, whose faces of any codimension are again hypercubes.      The key property  that we will use is that  topologically 
a hypercube of any dimension is a ball, and its boundary is a sphere.   In the following,  $\epsilon$ is an arbitrary,
sufficiently small parameter.  If it  is necessary to rescale $\epsilon$ by a fixed constant (which will happen finitely many times in the following), we do not
keep track of these constants and just call the new parameter $\epsilon$.   When we say ``$A$ is of order $\epsilon$,'' we always mean that $A$ and its field strength $F$ are of order $\epsilon$; this is not
always stated explicitly.

We will  use  some basic facts about the Dirichlet problem for Yang-Mills theory.   Before considering Yang-Mills theory, let us recall the analogous  and more elementary facts about Maxwell's equations.  
Given any abelian gauge field $A$ on the boundary of a ball $B$ of dimension $\geq 3$, 
 $A$ can be extended to a solution of Maxwell's equations on $B$ in a way that is unique up to a gauge transformation trivial on $\partial B$; 
 moreover, there is a unique gauge transformation on $B$ that is trivial on $\partial B$
 and  puts $A$ in Landau gauge  $\nabla^i A_i=0$ ($\nabla_i$ is the covariant derivative defined using the metric on $B$, not necessarily assumed to be flat).\footnote{Both main steps here can be viewed as the outcome of minimization problems.   Solving Maxwell's equations with given boundary values amounts
 to minimizing the Maxwell action $\int_B \d^dx\sqrt g F_{ij}F^{ij}$, and imposing the Landau gauge condition amounts to minimizing $\int_B \d^dx \sqrt g A_i A^i$ under gauge transformations that
 are trivial on the boundary (and, in $d=1$, are constrained to keep $A$ small).   The minima exist because the functionals in question are bounded below, and are unique because (after imposing
 smallness conditions in the low dimensional cases to avoid topological issues) these functionals are effectively quadratic functionals on connected spaces of fields.}   
  We call the gauge field satisfying these conditions the solution of the Landau gauge Dirichlet problem for $A$.   Uniqueness means that if we are
 given two Landau gauge solutions $A,A'$ of Maxwell's equations on $B$, then any gauge transformation between the boundary values of $A$ and of $A'$ extends uniquely to a gauge transformation
 from $A$ to $A'$ defined on all of $B$ .   These statements need some modification in dimension 2, where $\partial B$  is not simply connected, and in dimension 1, where $\partial B$
 is not connected.   In dimension 2, the same statements hold if the holonomy $U=\exp(\i \oint_{\partial B} A)$ is close enough to the identity, say $|U-1|<\epsilon$ for small $\epsilon$,
 and one stipulates that one wants the solution on $B$ to be small, with  $|A|, \,|F|<\epsilon$ everywhere (in units of the radius of $B$).   Now consider the case $d=1$.   The ball is then an interval, say the interval
 $0\leq x\leq a$, with metric $\d x^2$.   
To get a good analog of the Landau gauge Dirichlet problem in one dimension,
 one has to specify the desired value of $U=\exp(\i\int_B A) $, and this should be sufficiently close to 1, say $|U-1|<\epsilon$. One also has to stipulate that one wants $|A|$ to be of order $\epsilon$
 (in units of $1/a$).   Then the Landau gauge Dirichlet problem has a unique solution with $|A|<\epsilon/a$.
 Indeed, in one dimension Maxwell's equations are trivial, and  the Landau gauge condition $\nabla^i A_i=0$ says that $A$ should be constant. The unique solution of the Landau gauge Dirichlet
 problem is then $A=\frac{1}{\i a}\log U \,\d x$.  
 
 The Yang-Mills equations are analogous to Maxwell's equations, but nonlinear.   The nonlinearity, however, is small for weak fields.  All statements that we made about the Landau gauge Dirichlet problem
 for Maxwell's equations also hold for the Yang-Mills equations provided that,  to make the effects of the nonlinearity sufficiently small, we  stipulate that the assumed boundary values of  $|A|$ and $|F|$ are small, and that we want a bulk solution with $|A|$ and $|F|$ small.   In all cases, ``smallness''
 is measured in units of the radius of the ball or of its boundary.   The basis for the claim is the following.\footnote{For
 a rigorous discussion of a closely related Yang-Mills Dirichlet problem in four dimensions, see \cite{Marini}.}   The Yang-Mills equations plus the Landau gauge condition
 (and similarly their linearization in Maxwell theory) are a system of elliptic partial differential equations.
 In general, if a linear elliptic equation has a unique solution for all boundary data, then after perturbing it with a sufficiently small nonlinearity -- which we achieve here by requiring $|A|, |F|<\epsilon$ --
 the resulting nonlinear equation has a nearby solution that is unique and nondegenerate.  (Nondegeneracy means that in expanding around this solution of the nonlinear equation, there 
 are no zero-modes.)  
In our context, the conclusion
 is that given a Yang-Mills field on $\partial B$ with $|A|,|F|<\epsilon$, there is a unique Landau gauge solution of the Yang-Mills equations on $B$ with the given boundary
values and $|A|,|F|<\epsilon$.  
A variant of the uniqueness statement is the following:   if $A$, $A'$ are two Landau gauge solutions on $B$, both of order $\epsilon$
 and with field strength of order $\epsilon$, then any gauge transformation on $\partial B$ from the boundary values of $A$ to the boundary values of $A'$ extends uniquely to a gauge transformation from $A$ to $A'$ on $B$.   Note that controlling the nonlinearity in Yang-Mills theory has required the same sort of smallness condition on $A,F$ that was needed in Maxwell theory for small $d$ to deal with topological issues.
 
 Now we turn to lattice gauge theory. An admissible  lattice gauge field is  a $G$-valued function $U_{\langle xy\rangle}$ defined for nearest neighbor pairs of lattice points, satisfying $U_{\la yx\ra}=U^{-1}_{\la xy \ra}$
and such that  the plaquette holonomies $U_P$ satisfy $|U_P-1|<\epsilon$, for all $P$, with some $\epsilon$ that is sufficiently small for the following argument.     Given this data, we wish to construct a continuum
gauge field $A$ on $T^4$ that interpolates the lattice gauge field, in the sense that for each nearest neighbor pair $xy$, the parallel transport by $A$ along the link from $x$ to $y$ gives precisely
$U_{\langle xy\rangle}$.    The interpolation procedure will be gauge-invariant, in the sense that if two lattice gauge fields are gauge-equivalent, then their interpolations will also be
gauge-equivalent.
Moreover, the Yang-Mills field strength $F=\d A+A\wedge A$ of the interpolated gauge field
will be of order $\epsilon$ in lattice units; that is, if  the lattice spacing is $a$, then $F\sim \epsilon/a^2$.   

An  elementary fact that we will use  is the existence of a logarithm for a group element that is close to the identity.   In general, for a compact Lie group, there is no continuous map
$U\to 
\log U$ from the group to its Lie algebra satisfying $U=e^{\log U}$.   But in a neighborhood of the identity, there is a unique such map with $\log 1 =0$. We already used the existence of such a logarithm
in Maxwell theory for $d=1$.

\begin{figure}
 \begin{center}
   \includegraphics[width=1.5in]{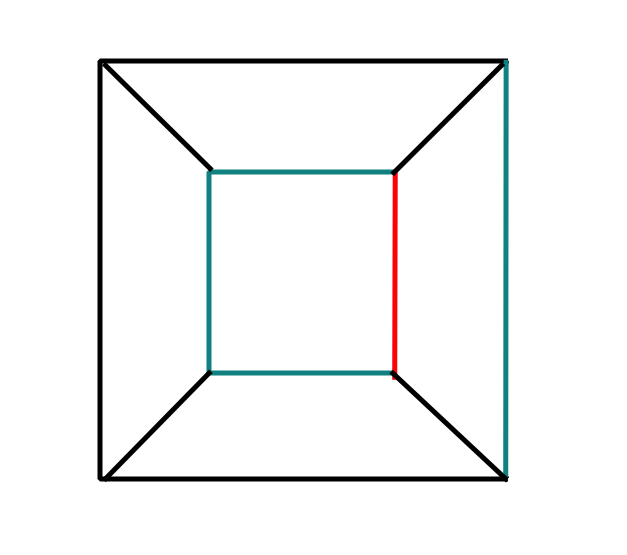}
 \end{center}
\caption{\footnotesize  The boundary of a three-dimensional cube has been projected to a plane, for visibility.   The 1-skeleton of the boundary consists of 12 links.   By a lattice gauge transformation,
up to 7 of the link variables can be set to 1, for example, the ones associated to the seven links shown in black, which make up a ``maximal tree.''  In that gauge, the link variables associated to the
four green links 
are plaquette holonomies and satisfy  $|U-1|<\epsilon$, while the remaining link variable in red
is the holonomy around the boundary of the union of two plaquettes (the outer one and one on the right), and so satisfies $|U-1|<2\epsilon$. For
a similar four-dimensional construction, start with a  ``bottom'' cube and a ``top'' one and connect each of the  8 corners
of the bottom cube to the corresponding corner of the top cube by a ``vertical'' link.   Gauge to the identity  7 link variables  in the bottom cube, as in the picture,  and the 8 vertical ones.    In this gauge, each
 remaining link variable is the holonomy around a region that encloses at most three plaquettes, so  all satisfy $|U-1|<3\epsilon$.   \label{four}}
\end{figure}

In lattice gauge theory on a hypercubic lattice,  the holonomy around a single plaquette is a product $U_P=U_1 U_2 U_3 U_4$ of four link variables.   It is possible to set three of them, say $U_1,U_2,U_3$,  to 1 by a gauge transformation,
and once this is done, one has $U_4=U_P$.   In the case of an admissible lattice gauge field, $|U_P-1|<\epsilon$, and therefore in the gauge stated, all of the link variables of the plaquette satisfy 
$|U-1|<\epsilon$.   Similarly, the boundary of a cube contains 12 edges, and (see fig. \ref{four}) it is possible to set 7 of them to the identity by a gauge transformation; once this is done, each remaining
link variable is the  holonomy around a lattice path that bounds the union of at most two plaquettes. So each link variable in this gauge obeys $|U-1|<2\epsilon$.   Finally, in a four-dimensional hypercube (as
explained in the caption to fig. \ref{four}),
it is possible to gauge 15 of its  edges to 1, and each remaining one is then the holonomy around a path that bounds at most three plaquettes. So in this gauge,  each link variable of the hypercube
satisfies $|U-1|<3\epsilon$.  (The generalization to a $k$-dimensional hypercube is $|U-1|<(k-1)\epsilon$.)

We will start by trying to define $A$ on a single four-dimensional hypercube $C_4$.   First we go to the gauge just described in which all link variables satisfy $|U-1|<3\epsilon$.
This means that the link variables all have logarithms.   On an edge with link variable $U$, we define $A$ to be the constant\footnote{The factor $1/\i$ is needed if $A$ is defined
to be hermitian and holonomy is defined by $U=P\exp\i\int\A$, where $P$ represents path ordering.   If $A$ is defined to be antihermitian and $U=P\exp\int\A$, then $A=\frac{1}{a}\log U$.}
  $A=\frac{1}{\i a}\log U$.   In one dimension, just as in the abelian case,
the Landau gauge condition $\sum_i \nabla^i A_i=0$ says that $A$ should be constant, and the Yang-Mills equations are vacuous, so we can describe the choice $A=\frac{1}{\i a}\log U$ as the unique Landau
gauge solution of the Yang-Mills equations that is of order $\epsilon$ and whose parallel transport along the edge agrees with $U$.   Next let $C_2$ be a two-dimensional face of $C_4$.   We have
already defined $A$ on $\partial C_2$, in such a way that it is of order $\epsilon$.   So solving the Landau gauge Dirichlet problem on $C_2$, we get a unique extension of $A$ over $C_2$ that is of
order $\epsilon$.
After doing this for every two-dimensional face of $C_4$, we consider in the same way a three-dimensional face $C_3$.   At this point, we have defined $A$ on $\partial C_3$ and it is of order $\epsilon$.
So solving the Landau gauge Dirichlet problem again, we extend $A$ over $C_3$.    After doing this for each three-dimensional face of $C_4$, we have defined $A$ on $\partial C_4$, and it is of
order $\epsilon$.   So we solve the Landau gauge Dirichlet problem one more time and extend $A$ over $C_4$, again of order $\epsilon$.

Now let us discuss how to combine the results of this construction on  hypercubes $C$ and $C'$ that share a common 
 face $C\cap C'$.   The starting point of the whole construction was, of course, an admissible  lattice gauge field
defined on the whole lattice, with link variables $U_{\la xy\ra}$ for every link $xy$.   Then on $C$, we made a lattice gauge transformation $\phi_C$ -- defined only on the lattice sites -- to ensure
that all link variables on $C$ satisfy $|U-1|<3\epsilon$, and this was the starting point for the construction of an interpolating gauge field $A_C$ on $C$.   Similarly, on $C'$, we made a lattice gauge
transformation $\phi_{C'}$ to a gauge in which  all link variables on $C'$ obey $|U-1|<3\epsilon$, and this led to the construction of an interpolating gauge field $A_{C'}$ on $C'$.   In general the
gauge transformations $\phi_C$ and $\phi_{C'}$ are not equal on $C\cap C'$ (and it is not possible through additional choices to make them equal for all adjacent pairs $C,C'$).  However,
we can construct on $C\cap C'$  a uniquely determined gauge transformation from $A_C$ to $A_{C'}$.   First of all, since the lattice data that were used as input to the interpolation on $C$ and on $C'$
were obtained by lattice gauge transformations $\phi_C$ and $\phi_{C'}$ from the same initial family $U_{\la xy\ra}$, it follows that, on $C\cap C'$,  the lattice data used on $C'$ are related to the lattice 
data used on $C$
by the gauge transformation $\phi_{C'}\phi_C^{-1}$.   Second, the subsequent steps in which starting with lattice data on $C$ or on $C'$, we constructed the interpolating gauge fields $A_C$ and $A_{C'}$
first on links, then on squares, then on cubes, and finally on all of $C$ or $C'$, were made at each stage by solving a Landau gauge Dirichlet problem.   The uniqueness of the solution of the Landau gauge Dirichlet problem means that any gauge equivalence between boundary values extends uniquely to a gauge equivalence between bulk solutions. Applying this reasoning first on links, then on squares
and so on, we learn that  there is a uniquely determined
gauge transformation $\Phi_{C',C}$ from $A_C$ to $A_{C'}$ that agrees with $\phi_{C'}\phi_C^{-1}$ on the lattice sites.   In general, $\Phi_{C',C}$ does not factorize as 
$\psi_{C'} \psi_C^{-1}$ where $\psi_{C'}$ and $\psi_C$ respectively extend $\phi_{C'}$ and $\phi_C$ over all of $\partial C'$ and $\partial C$.   If that is the case, the bundle on $T^4$ constructed
presently is topologically trivial.   Note that this argument constructing $\Phi_{C',C}$
holds  equally regardless of the dimension of $C\cap C'$ (which may be 0,1,2, or 3, but if it is of dimension 0, there was no nontrivial statement to prove).

Next  consider three hypercubes $C,C',C''$ with non-empty joint intersection. 
On $C\cap C'\cap C''$, we can define the following two  gauge transformations from
$A_C$ to $A_{C''}$, namely $\Phi(C'',C)$ and $\Phi(C'',C')\Phi(C',C)$.   We want to know that these are equal,
\be\label{needtoknow} \Phi(C'',C)=\Phi(C'',C')\Phi(C',C). \ee
This is true on lattice sites, since on lattice sites $\Phi(C'',C)=\phi_{C''}\phi_C^{-1}$. $\Phi(C'',C')=\phi_{C''}\phi_{C'}^{-1}$, $\Phi(C',C)=\phi_{C'}\phi_C^{-1}$.  The uniqueness part of the Landau gauge Dirichlet
problem, applied  just as we did to define $\Phi(C',C)$, then shows that the relation (\ref{needtoknow}) must also hold on all of $C\cap C'\cap C''$.    Again, this argument holds regardless of the dimension
of $C\cap C'\cap C''$ (which may be 0, 1, or 2).

Each $A_C$ or $A_{C'}$ can be viewed as a connection on a trivial $G$-bundle over $C$ or $C'$.   The gauge transformation $\Phi(C',C)$ can be used to glue together the bundles over $C$ and $C'$,
and their corresponding connections $A_C$ and $A_{C'}$, to build a $G$-bundle with connection over the union $C\cup C'$.   The relation (\ref{needtoknow})   means that no inconsistency is encountered
in further gluing of hypercubes, so we can glue together the data on all the hypercubes comprising $T^4$ to build a $G$-bundle with connection $A$ over all of $T^4$.    The field strength $F$  of this connection
is everywhere of order $\epsilon$ in lattice units.  The bundle constructed in this way may be topologically nontrivial (if the bundle is nontrivial, a global gauge choice does not exist; even if
it is trivial, there may be no global gauge choice with $A$ everywhere small, though there is such a choice in any one hypercube or small set of hypercubes).   
In fact, some of the steps in this argument are similar to reasoning used by L\"uscher \cite{Luscher3} to define the topological charge of an admissible lattice gauge field.
To make it obvious that there is no general restriction on the topology of the bundle constructed in this way, observe that starting with any $G$-bundle over $T^4$, any connection $A$ on this bundle,
if restricted to a sufficiently fine lattice, becomes an admissible lattice gauge field; then the interpolating procedure will simply reconstruct the $G$-bundle that we started with, up to isomorphism, with a connection
close to the original one. 

The inductive nature of this construction hopefully makes it clear that the same construction applies for a hypercubic lattice of any dimension.
The continuum gauge field produced by this construction has the property that its holonomy on a  given link agrees precisely with the corresponding lattice link variable, and
 its restriction to any given $k$-cell depends only on the lattice link variables of that cell.
It is smooth in the interior of each four-dimensional hypercube.   Across boundaries of hypercubes, the interpolated gauge field is continuous but its derivative is not continuous.  So the Yang-Mills
field strength, while everywhere of order $\epsilon$ in lattice units, is discontinuous across boundaries of hypercubes.  This actually should not interfere with the use of this interpolated
gauge field to define a coupling of lattice gauge theory to chiral fermions, since the formula $|\det\,\slashed{D}_W^+|\exp(-\i\pi \eta(\slashed{D}_Y))$ makes sense with this kind of gauge field.
The behavior of the interpolation that we have described  across boundaries of hypercubes is actually similar to the typical behavior of lattice gauge fields at the lattice scale.    However,  as noted in \cite{Hsu2},
a numerical evaluation of $|\det\,\slashed{D}_W^+|\exp(-\i\pi \eta(\slashed{D}_Y))$
is likely to be more efficient if chiral fermions are coupled to a smooth gauge field.   

 There are several ways to achieve smoothness.
A simple smoothing procedure is to act on the interpolated $A$ with the Yang-Mills gradient flow for a short time.  To define this flow, let $s$ be a new ``time'' coordinate,
and let the four-dimensional gauge field $A$ evolve by
\be\label{gflow} \frac{\d A_i(x)}{\d s}= \sum_j D_j F_{ji}(x). \ee
Starting at $s=0$ with the continuum interpolation $A$ that we already constructed, this will produce a smooth continuum gauge field for any $s>0$.   Gradient flow for a time $s_0$ will
smooth out $A$ on a length scale $\sqrt{s_0}$, so picking $s_0\sim a^2$ will suppress modes of $A$ with wavelength smaller than the lattice spacing $a$.   This  gives a smooth continuum
gauge field in the same topological class as $A$; its link holonomies agree with $U_{\la xy\ra}$ up to small corrections (but not exactly), and it is quasilocal in the sense that dependence on links
that are $n$ lattice units away is suppressed by $e^{-n^2 a^2/s_0}$. All this is sufficient for coupling to continuum fermions.\footnote{Craig's
interpolation for $SU(N)$ \cite{Craig} has the following properties.   It is smooth; it is strictly local in the sense that $A(x)$ does not depend at all on lattice links more than a fixed distance away;
and its link holonomies do not precisely coincide with $U_{\la xy\ra}$.}   

Anthropic's Claude Opus 5.5 suggested a  likely variant that gives a smooth interpolation with link holonomies  that equal precisely the corresponding lattice link variables 
and such that the interpolated gauge field in a hypercube depend only on the link variables on the boundary of that hypercube.  Pick coordinates on $\R^4$ so that the lattice consists of the points with
integer coordinates.   Fix a small $\delta>0$, and let $s(t)$ be a smooth nondecreasing function, with $s(t+1)=s(t)+1$, that is equal to $n$ for $|t-n|<\delta$, $n\in\Z$, and maps each 
interval $[n,n+1]$ onto itself. The map $\Phi(x)=(s(x_1),\dots,s(x_4))$ maps every cell of the lattice onto itself; it collapses the $\delta$-neighborhood of the boundary of the cell 
onto the boundary, and maps the rest of the cell onto its interior. After defining $A$ on the cells of dimension $k$, and before going on to dimension $k+1$ by solving a Dirichlet problem, 
replace $A$ on each $k$-cell by $\Phi^*A$. This squeezes the gauge field on each cell into its interior and, near each face, replaces it by its value on the face, continued without change 
in the normal directions. Since $\Phi$ acts on each link only by reparametrization, the link holonomies are unchanged. Gauge fields on adjacent cells now match smoothly across their common faces.
The end result is a smooth continuum interpolation  procedure that is gauge-invariant, preserves all lattice symmetries, reproduces the link variables exactly, and on each cell depends only on the lattice data of that cell.

There are many possible variants of this construction, some  with possibly desirable properties.
For example,  Opus 5.5 also  described a variant in which the inductive step is made by a sort of linear interpolation
rather than by solving a Dirichlet problem. 

Continuum interpolation is also available, with similar arguments, if lattice gauge theory is formulated on a general smooth 
$d$-manifold $W$, for any $d$.   To do lattice gauge theory on $W$, we want not just a lattice embedded in $W$
but a sequence of  increasingly fine lattices in $W$ on which lattice gauge theory   can be expected to converge to a continuum limit.  
For this, we want the top dimensional cells in the refined lattices to have ``bounded flatness,'' that is, as the lattice is refined, one wants these cells to remain
roughly isotropic, rather than becoming highly flattened in some directions relative to others, which would make it difficult to see a continuum limit of lattice gauge theory.\footnote{A procedure that does not work well is barycentric subdivision, in which one starts with a triangulation and at each stage adds a new vertex in each simplex and subdivides. The simplices become very flat.}
 One way to proceed (in any dimension) is as follows, following
Freudenthal \cite{Freudenthal} (the title of the paper can be translated ``Simplicial Decompositions of Bounded Flatness'').  
Smooth manifolds of any dimension can always be triangulated \cite{Cairns,Whitehead}; start with any triangulation of $W$.   In a triangulation, the basic cells are simplices (a $k$-simplex is the $k$-dimensional generalization
of a tetrahedron, with $k+1$ vertices), as opposed to the hypercubes in our previous discussion.\footnote{Simplices are simpler than hypercubes in one respect. After gauging to 1 a maximal
number of link variables in a simplex, the remaining ones are all plaquette  holonomies.    Thus on the 1-skeleton of a simplex, an admissible lattice connection can always be put in a gauge
with $|U-1|<\epsilon$ for all links, while for hypercubes of dimension $k$, $\epsilon$ must be replaced by $(k-1)\epsilon$.}  Order the vertices in the triangulation.    Then in particular the vertices $v_0,v_1,\cdots, v_k$ of any
given $k$-simplex are ordered.  The simplex can be identified with a region $\{1\geq x_1\geq x_2\geq\cdots \geq x_k\geq 0\}$ in $\R^k$; the vertices are points at which all but one of these inequalities are saturated
and are ordered accordingly.
Picking any integer $n>1$, subdivide the simplex into $n^k$ smaller ones by cutting along the hyperplanes $x_i=m/n$ and $x_i-x_j=m/n$, with integers $0<m<n$.    This divides the original
simplex into $n^k$ smaller ones. (This decomposition of a $k$-simplex restricts along its boundary to the decomposition of its faces made in the same way.)   As $n\to\infty$, the simplices remain of bounded flatness, because regardless of $n$, the simplices that arise in this process have only $k!/2$ possible shapes
\cite{EG}.   The vertices in each simplex of the refined lattice inherit an ordering from the ordering of the vertices in the original lattice, so the process of subdivision can be repeated.
Subdividing in this fashion with $n=n_1n_2$ is equivalent to first subdividing with $n_1$ and then with $n_2$, so for example, to subdivide with $n=2^b$ can be accomplished with $b$
steps in each of which one subdivides by a factor of 2.  So this process of subdivision may be amenable to a renormalization group approach.
To do lattice gauge theory in this setting, one can start with the Wilson action defined by a sum over plaquettes $I=\sum_P w_P {\mathrm{Re}}\Tr (1-U_P)$, but now with different weights $w_P>0$ for different plaquettes.
The weights would be adjusted so that, expanding around a configuration with all link variables equal to 1,
 the quadratic part of the action matches the desired $\frac{1}{4 g_{\mathrm{YM}}^2} \int \d^4x \sqrt g g^{ij} g^{kl} \Tr\,F_{ik} F_{jl}$, with some desired metric $g_{ij}$ of $W$ and some Yang-Mills coupling 
 $g_{\mathrm{YM}}$.    Near the continuum limit, one takes $g_{\mathrm{YM}}$ to zero in the usual way; some further tuning of the lattice weights will be required to avoid generating non-isotropic
 dimension four couplings by quantum corrections.
 
 There is no problem  with continuum interpolation in this setup.   For example, the approach based on the solution of the Dirichlet problem carries over
 without any essential change.  The cells in the lattice are now simplices, rather than the hypercubes of the previous discussion.   But as a simplex is topologically a ball, with its boundary a sphere,
 this does not require any essential change in the previous explanation based on the Dirichlet problem.

\section{Can This Regularization Be Implemented on a Computer?}\label{effective}

Finally, we will address this question: can the regularization of chiral gauge theories that we have discussed, with lattice gauge fields interpolated over the continuum and coupled to continuum chiral
fermions, be implemented on a computer?

There will certainly be no straightforward  lattice Monte Carlo implementation, because the chiral fermion path integral, even when anomaly-free, lacks the positivity needed in Monte Carlo procedures.
So an implementation, at least with known methods, may be impracticably slow.   But is such an implementation possible in principle?
In lattice gauge theory with a finite volume and a positive lattice spacing, and therefore only finitely many link variables,
the lattice path integral is an integral over a compact space of finite dimension.
  So if the chiral fermion integral  can be  effectively computed, then the whole lattice path integral can be effectively computed as well,
in principle.   

So we have to discuss whether one can effectively interpolate an admissible lattice gauge field over the continuum, extend it over some $Y$, and then compute
\be\label{welzox} |\det\,D_W^+| \exp(-\i\pi\eta(\D_Y))\ee
(and its generalization in instanton sectors, which raises similar issues).
The interpolation is certainly not a problem; with the procedure described in section \ref{interpolation}, this amounts to solving finitely many Landau gauge Dirichlet problems, which
certainly can be solved numerically.   The extension over $Y$ is also not an issue.   For example, for  $T^4= T^3\times S^1$, we can take $Y=T^3\times D$, where $D$ is a disc of
boundary $S^1$.   With a connected, simply-connected group $G$, a  topologically trivial gauge field $A$  can be extended over $Y$ and such an extension can be effectively found.  Indeed, any
path from $A=0$ to a given $A$ can be used to construct an extension of $A$ over $T^3\times D$. (To do this, use the path to construct a gauge field on $S^1\times I$, where $I$ is an interval,
with $A=0$ in a neighborhood of one end and the desired $A$ at the other end; then collapse the end with $A=0$ to build a disc with a gauge field.)  Variants of this are possible in topologically non-trivial sectors.   No particularly nice procedure
is needed for the extension because the result will anyway be independent of $Y$.

The question is then whether one can effectively approximate $|\det\,\D_W^+|$ and $\exp(-\i\pi\eta_Y)$ on a computer.  Will the fermion doubling problem come back when we try to do so?
In fact, there is no difficulty.     Although we would face a fermion doubling problem if we try to make sense of $\det\,\D_W^+$ directly, there is no problem to compute its absolute value $|\det\,\D_W^+|$.
That is the square root of the determinant of the self-adjoint operator $\D_W^{+\dagger}\D_W^+$.   This operator is Laplace-like, and there is no difficulty to discretize it and compute its spectrum.
Now let us discuss the problem of computing the eta-invariant of the Dirac operator $\D_Y$, a self-adjoint operator defined with APS boundary conditions.
   This eta-invariant of course can be extracted from the spectrum of $\D_Y$.   If we try to compute the spectrum of
$\D_Y$ directly, for example by a lattice regularization, we may indeed recover a version of the lattice doubling problem.   However, there is no need to do that.   We can compute the spectrum
of the self-adjoint operator $\D_Y^2$ (defined with a boundary condition that is inherited from the APS boundary condition of $\D_Y$).
   This is a Laplace-like operator with no doubling problem.   Generically, the eigenvalues of $\D_Y^2$ are nondegenerate.  If $\psi$ is a eigenvector of $\D_Y^2$
with nondegenerate eigenvalue $\lambda$, then it is an eigenstate of $\D_Y$ with eigenvalue $\pm \sqrt \lambda$, and we can find the sign by computing the expectation
value $(\psi,\D_Y\psi)$.   If exceptionally $\D_Y^2$ has an $s$-dimensional space of eigenvectors with the same eigenvalue $\lambda$, we evaluate $\D_Y$ is a matrix
in this space and learn the multiplicity of states with eigenvalue $+\sqrt \lambda$ or $-\sqrt\lambda$.   The APS boundary condition that was part of the input here can be described
just in terms of the nonchiral Dirac operator $\D_W$, so it too can be approximated numerically without running into a doubling problem.   So in short, there is no fermion doubling
problem in numerical study of the eta-invariant $\eta(\D_Y)$.

\vskip1cm
 \noindent {\it {Acknowledgements}}  
  Research supported in part by NSF Grant PHY-2514611.  I thank T. Hartman for pointing out Craig's paper, and K. Yonekura for helpful comments on the manuscript.
   Assistance on many details was provided by Claude Opus 5.5.
 \bibliographystyle{unsrt}

\end{document}